\documentclass{ws-ijmpb}
\pdfoutput=1
\usepackage{url}
\usepackage{amsmath,amssymb}
\usepackage[super,compress]{cite}
\begin{document}

\markboth{Sudip Chakravarty and Chen-Hsuan Hsu}
{SKYRMIONS IN A DENSITY WAVE STATE...}

%
\catchline{}{}{}{}{}
%

\title{SKYRMIONS IN A DENSITY-WAVE STATE: A MECHANISM FOR CHIRAL SUPERCONDUCTIVITY}

\author{Sudip Chakravarty}

\address{Department of Physics and Astronomy, University of California Los Angeles\\
Los Angeles, California 90095,
USA\\
sudip@physics.ucla.edu}

\author{Chen-Hsuan Hsu}

\address{RIKEN Center for Emergent Matter Science (CEMS)\\
Wako, Saitama 351-0198, Japan\\
chenhsuan.hsu@gmail.com}

\maketitle

\begin{history}
\received{Day Month Year}
\revised{Day Month Year}
\end{history}

\begin{abstract}
Broken symmetry states characterizing density waves of higher angular momentum in correlated electronic systems are intriguing objects. In the scheme of characterization  by angular momentum, conventional charge and spin density waves correspond to zero angular momentum. Here we explore a class of exotic density wave states that have   topological properties observed in recently discovered topological insulators. These rich topological density wave states deserve closer attention in not only high temperature superconductors but in other correlated electron states, as in heavy fermions, of which an explicit example will be discussed.
The state discussed has non-trivial charge $2e$ skyrmionic spin texture. These skyrmions can condense into a charged superfluid. Alternately, they can fractionalize into merons and anti-merons. The fractionalized particles that  are confined in skyrmions in the insulating phase, can emerge at a deconfined quantum critical point, which separates the insulating and the superconducting phases. These fractional particles form a two-component spin-singlet chiral $(d_{x^2-y^2}\pm id_{xy})$ wave superconducting state that breaks time reversal symmetry. Possible connections of this exotic order to the superconducting state in the heavy-fermion material URu$_2$Si$_2$ are suggested. The direct evidence of such a chiral superconducting state is polar Kerr effect that was observed recently.  
\end{abstract}


\section{Introduction}

In this review we address novel role that skyrmions~\cite{Skyrme:1962} and their fractionalized avatars, merons and anti-merons, can play in certain correlated electron systems. This review  focuses on our own work~\cite{Hsu:2011,Hsu:2013,Hsu:2014} except where contributions by other authors have provided  ingredients.  The exploration of skyrmions in the density wave states of the kind discussed here has,  to our knowledge, not been discussed in places other than in our own work. One of the key features is that it can predict a chiral $d$-wave superconducting state  that breaks time reversal symmetry (TRS), which appears to have been directly observed in recent polar Kerr effect (PKE) measurements by the Stanford group~\cite{Schemm:2015} in a heavy fermion material: $\mathrm{URu_{2}Si_{2}}$. Broken TRS  is a necessary condition for  a nonzero  PKE.~\cite{Halperin:1992} 

The  order parameter in $\mathrm{URu_{2}Si_{2}}$ (URS), a heavy-fermion material, below the so called hidden-order (HO) transition at $T_{\textrm{HO}}=17.5$~K is unknown despite its discovery over a quarter century ago; see Ref.~\refcite{Mydosh:2011} and Ref.~\refcite{Mydosh:2014} and references therein. Within this phase lies a much less explored unconventional  superconducting state with a transition temperature $T_{c}\sim 1.5$~K.~\cite{Palstra:1985,Maple:1986,Schlabitz:1986,deVisser:1986,Mydosh:2011}  It stands to reason that there must be an intimate relation between the two. While numerous theoretical models have been proposed to explain the HO phase,~\cite{Mydosh:2011,Mydosh:2014} there are very few attempts to explain the mechanism of the unconventional superconductivity. It is our central interest to explore the connection between the two states to provide a skyrmionic mechanism for the unconventional superconducting state, which arises from an intriguing  density wave state, termed mixed singlet-triplet $d$-density wave (st-DDW).~\cite{Hsu:2011,Hsu:2013,Hsu:2014}  This state  has no net charge  or spin modulations and does not break TRS. It does  have  topological order with quantized spin Hall effect.~\cite{Hsu:2011} Thus, it is naturally impervious to common experimental probes and can be aptly described as an hidden order state. Determination of the  density wave state posited here   may be possible through two-magnon Raman scattering, nuclear quadrupolar resonance, or  the skyrmions themselves.    In a more general context, our work reflects the rich possibilities of emergent behavior in condensed matter systems. Attempts were made to describe the HO in terms of the triplet $d$-density state (addressed below) to explain the observed anisotropic magnetic susceptibility.~\cite{Fujimoto:2011,Okazaki:2011,Shibauchi:2012}  While  this is an interesting idea,  so far it has not been able to provide a mechanism for superconductivity, which must be related to the HO state.

An early attempt, with some family resemblance to the skyrmionnc mechanism discussed here, is the non-BCS mechanism   of superconductivity   suggested  by Wiegmann,~\cite{Wiegmann:1999} as an extension of Fr\"ohlich mechanism in one dimension to higher dimensions. The crux  was the concept of spectral flow. Consider gapped fermions in a static potential. Assume that the chemical potential $\mu$ lies in the gap. When we  change the potential adiabatically, the  energy levels of the fermions move around. Typically the levels cannot cross $\mu$, but there are potentials  such that  an adiabatic and  smooth variation  creates even number of unoccupied states below  $\mu$, or forces some occupied levels to cross this level. For this spectral flow to occur, the variation of the potential, a soliton, must necessarily be topological. The index theorem then relates the topological charge of the soliton to the number of levels crossed. This phenomenon produces a compressible liquid, a superconductor.

More recently,  several interesting papers have  led to discussions of superconductivity  in single and bilayer graphenes. Grover and Senthil~\cite{Grover:2008} have  provided a mechanism in which  electrons hopping on a honeycomb lattice can lead to a charge-$2e$ skyrmionic condensate, possibly relevant to  single layer graphene. As to  bilayer graphene,  a charge-$4e$ skyrmionic condensate  has been suggested by Lu and Herbut~\cite{Lu:2012} and Moon.~\cite{Moon:2012} See also the earlier  work in Ref.~\citenum{Sondhi:1993} of charge-$e$ skyrmions in a quantum Hall ferromagnet.

 The starting point is layered condensed matter systems with weak interplanar tunneling. For realistic application to $\mathrm{URu_{2}Si_{2}}$ these planes are along the body diagonals of a body-centered tetragonal crystal containing the active $\mathrm{U}$ atoms. This obscures the main features of the skyrmionic mechanism greatly. In addition, both crystal field effect and spin-orbit coupling must be taken into account, as was presented in the previous papers~\cite{Hsu:2011,Hsu:2013,Hsu:2014}, although, in the end, they turn out not to be of primary importance for the mechanism of superconductivity discussed here. Also, the nesting vector appropriate for this material should be $(0,0,2\pi/c)$, where $c$ is the height of the  unit cell.~\cite{Mydosh:2011} In order to reveal the principal aspects,  in the present paper we shall simplify by considering the weakly coupled layers to be in the $XY$-planes and the nesting vectors to be $(\pi/a,\pi/a,0)$ and its symmetry complements, where $a$ is the  spacing of the square planar lattice, which we shall set to unity. Only a minimal band structure involving nearest and next-nearest neighbor hopping will be incorporated. We shall eschew all other complicating details that can be found in our published papers.~\cite{Hsu:2011,Hsu:2013,Hsu:2014}

There are two  points that are crucial to our work. The first is rather simple: in the density wave state considered here, there are also Goldstone modes that can be easily seen by integrating out the fermions resulting in a 
 non-linear $\sigma$-model involving a unit vector $\hat{N}$, the form of which is entirely determined by symmetry and is given below. The Goldstones  are spectators to charge-$2e$ skyrmions that can possibly Bose condense, or their fractionalized avatars that lead to a paired BCS state at $T=0$. 
 At finite temperatures, however, they could lead to interesting behavior.~\cite{Chakravarty:1989} The second point is more subtle: we  assume that the hedgehog configurations are absent. This would require that the energy of the skyrmions be smaller than the  single electron gap, a question that is likely to be model dependent. If this assumption is correct,  the transition from  the st-DDW state (discussed in Section 2)  to the superconducting state will correspond to a deconfined quantum critical point at $T=0$, which otherwise would be a first order transition, as in Landau theory.~\cite{Senthil:2004,Senthil:2004b,Kuklov:2006,Grover:2008} In other words,   skyrmion number conservation  in each plane is crucial to the fractionalized mechanism explored here.

We beg the readers' patience in reading this manuscript. The path through this review  is as follows:
In Section 2 we begin with  the relatively ill-understood density wave states of higher angular angular momentum. A  case of crucial importance is what we term as the st-DDW. The state is a topological spin Hall insulator with quantized spin Hall conductance. In Section 3 we show that the quantum fluctuations from the mean field state define skyrmionic texture of charge $2e$, zero spin and zero angular momentum. In principle, this charge $2e$ skyrmions can Bose condense, akin to the seminal work of Wiegmann~\cite{Wiegmann:1999} (and even earlier by Fr\"ohlich). Although this in itself is interesting, we push the argument further in Section 4  and show that fractionalization of skyrmions can result in chiral superconductivity, which appears to have  been observed  $\mathrm{URu_{2}Si_{2}}$.  Section 5 then discusses the endpoint of our tortuous path. A pesky notational issue is the vector notation: if no confusions  arise we will set $\bf Q$ to be simply $Q$. Similarly, wherever possible  we shall set $\hbar=c=e=1$.

\section{Density Wave States of Non-Zero Angular Momentum}

Density wave states in correlated electron systems can be defined by the angular momentum quantum number and the fundamental nature of the condensates~\cite{Nayak:2000}. A superconductor is a condensate of Cooper pairs, that is, the condensation is in the particle-particle channel. Thus, the  antisymmetry of the wave function  provides a strict restriction on the spin function.  If the orbital function is symmetric, the spin function must be antisymmetric and vice versa. In contrast, the density wave states are condensates of bound pairs of electrons and holes. Because there are no requirements mandated by exchange between the two distinct particles, the orbital wave function cannot constrain the spin wave function. Although angular momentum is not a strict quantum number in a crystal, we will continue to use it as a metaphor---the proper classification is in terms of the symmetries of the point group.

\begin{figure}[htbp]
\begin{center}
\includegraphics[width=0.5\linewidth]{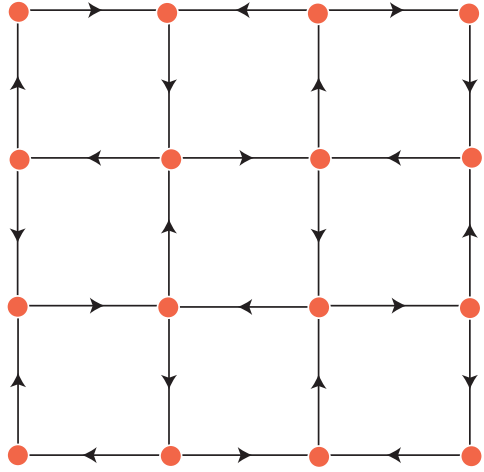}
\caption{Staggered pattern of charge currents reflecting DDW order in a square planar lattice.}
\label{fig:ddw-current}
\end{center}
\end{figure}

For a superconductor $\ell = 0, 1, 2, \ldots$ define $s$-wave (spin singlet), $p$-wave (spin triplet), and $d$-wave (spin singlet) condensates, etc. For a particle-hole condensate, that is, a density wave, $\ell =0$ comes in two varieties: the  spin singlet version defines  the familiar charge density wave (CDW) and the triplet version the spin density wave (SDW). The $\ell =1$ comes also in two versions and involves bond order. The case $\ell=2$, spin singlet, is not a wave of density at all, but corresponds to a staggered pattern of circulating charge currents, dubbed the $d$-density wave (DDW). The $\ell=2$, spin triplet,  corresponds to a staggered pattern of circulating spin currents. The two-fold commensurate DDW breaks translation,  time reversal, parity, and a rotation by $\pi/2$, while the product of any two symmetries is preserved. More specifically, the DDW order parameter, of period-2, is defined by 
\begin{equation}
\langle c^{\dagger}_{k+Q,\alpha}
c_{k,\beta}\rangle
= i\frac{\Phi_{Q}}{2}\,(\cos{k_x}-\cos{k_y})\, \delta_{\alpha,\beta},
\end{equation}
where ${Q} = (\pi, \pi)$. Note the similarity of the form factor with the $d$-wave superconductor (DSC), but the  factor of $i$ signifies the breaking of time reversal symmetry. The Kronecker $\delta_{\alpha,\beta}$ reflects the fact that the order parameter transforms as identity in the spin space, hence a singlet. As mentioned above,  the order parameter in the real space corresponds to a staggered pattern of circulating charge currents shown in Fig.~\ref{fig:ddw-current}.
It has been proposed~\cite{Chakravarty:2001,Laughlin:2014a,Laughlin:2014b} that the DDW gap is proportional to pseudogap $T^{*}$ in a cuprate superconductor. A triplet DDW order on the other hand is given by 
\begin{equation}
\langle c^{\dagger}_{k+Q,\alpha}
c_{k,\beta}\rangle
= i\frac{\Phi_{Q}}{2}(\cos{k_x}-\cos{k_y})\, \hat{N}\cdot \vec{\sigma}_{\alpha,\beta},
\end{equation}
which clearly transforms as a triplet under rotation in spin space; $\sigma$'s denote the standard Pauli matrices. In the real space this order parameter corresponds to circulating staggered spin currents. The unit vector $\hat{N}$ defines the direction of the spin. 

\begin{figure}[htbp]
\begin{center}
\includegraphics[width=0.5\linewidth]{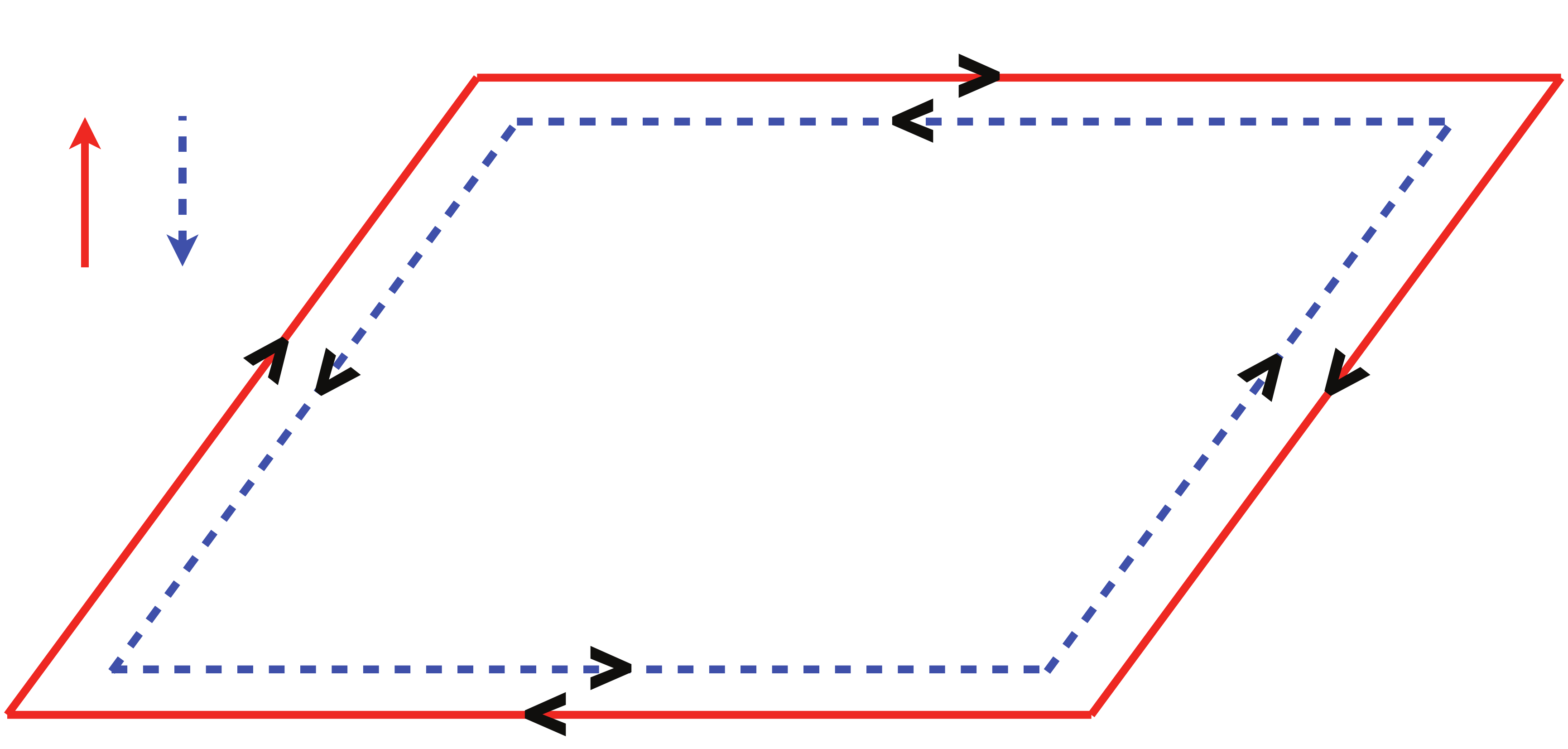}
\caption{Triplet $i \sigma d_{x^{2}-y^{2}}$ density wave in the absence of an external magnetic field.  The current pattern of each spin species on an elementary plaquette is shown.   The state is a semimetal. On the other
hand $i\sigma d_{x^{2}-y^{2}}+ d_{xy}$ can be fully gapped for a range of chemical potential. An example is shown in Fig.~\ref{band}: see Ref.~\citenum{Hsu:2011}}
\label{tDDW}
\end{center}
\end{figure}

\begin{figure}[htbp]
\begin{center}
\includegraphics[width=0.5\linewidth]{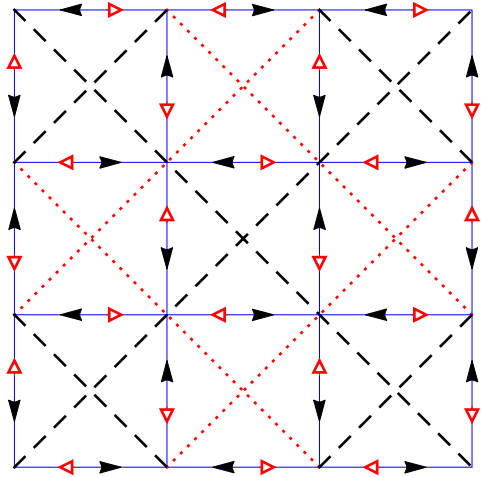}
\caption{st-DDW order on a square lattice. The solid (open) arrowheads indicate the current directions of up- (down-) spins due to the triplet $i\sigma d_{x^2-y^2}$ order. The solid lines indicate the nearest neighbor hopping. The dashed and dotted lines indicate different signs of modulations of the next-nearest neighbor hopping due to the singlet $d_{xy}$ order.}
\label{Fig:orders}
\end{center}
\end{figure}

It has been known that  triplet $i\sigma d_{x^{2}-y^{2}}$ order parameter corresponds to staggered circulating spin currents around a square plaquette.~\cite{Nersesyan:1991}. The oppositely aligned  spins circulate in opposite directions, as shown in Fig.~\ref{tDDW}. This reminds us of topological  insulators where oppositely aligned edge-spins travel in opposite directions. However, there is no topological protection because the bulk is not gapped, but is a semimetal instead. A more interesting  case is the st-DDW order parameter $(i \sigma d_{x^{2}-y^{2}}+ d_{xy})$, where $\sigma=\pm 1$ for up and down spins, with the quantization axis along $\hat{z}$.   In the momentum space, the order parameter will be
\begin{equation}
\langle c_{k+Q,\alpha}^{\dagger} c_{k,\beta}\rangle \propto [i(\vec{\sigma} \cdot \hat{N})_{\alpha\beta} W_k + \delta_{\alpha\beta} \Delta_k],
\end{equation}
and the form factors are
\begin{eqnarray}
W_k &\equiv& \frac{W_0}{2} (\cos \it{k_x} - \cos \it{k_y}),\\
\Delta_k &\equiv& \Delta_0  \sin k_x \sin k_y,
\end{eqnarray}
corresponding to the $d_{x^2-y^2}$ and $d_{xy}$ components, respectively. The current pattern is unchanged but the hopping matrix elements along the diagonals are modulated by $\Delta_{k}$,
 as shown in Fig.~\ref{Fig:orders}. 

The
singlet chiral $(i d_{x^{2}-y^{2}}+d_{xy})$ density wave, however,    breaks macroscopic time reversal symmetry and was employed to deduce possible polar Kerr effect and anomalous Nernst effect~\cite{Tewari:2008, Zhang:2009,Kotetes:2008,Kotetes:2010} in the pseudogap phase of the cuprates.

\subsection{Mean field theory of quantized spin Hall effect in st-DDW}
Unlike the semimetallic DDW, st-DDW has   a non-vanishing
 quantized  spin Hall effect for a range of values of the chemical potential. This is in fact a topological Mott insulator~\cite{Raghu:2008} because it is the electron-electron interaction that is necessary for it to be realized.  

\begin{figure}[htbp]
\begin{center}
\includegraphics[width=0.5\linewidth]{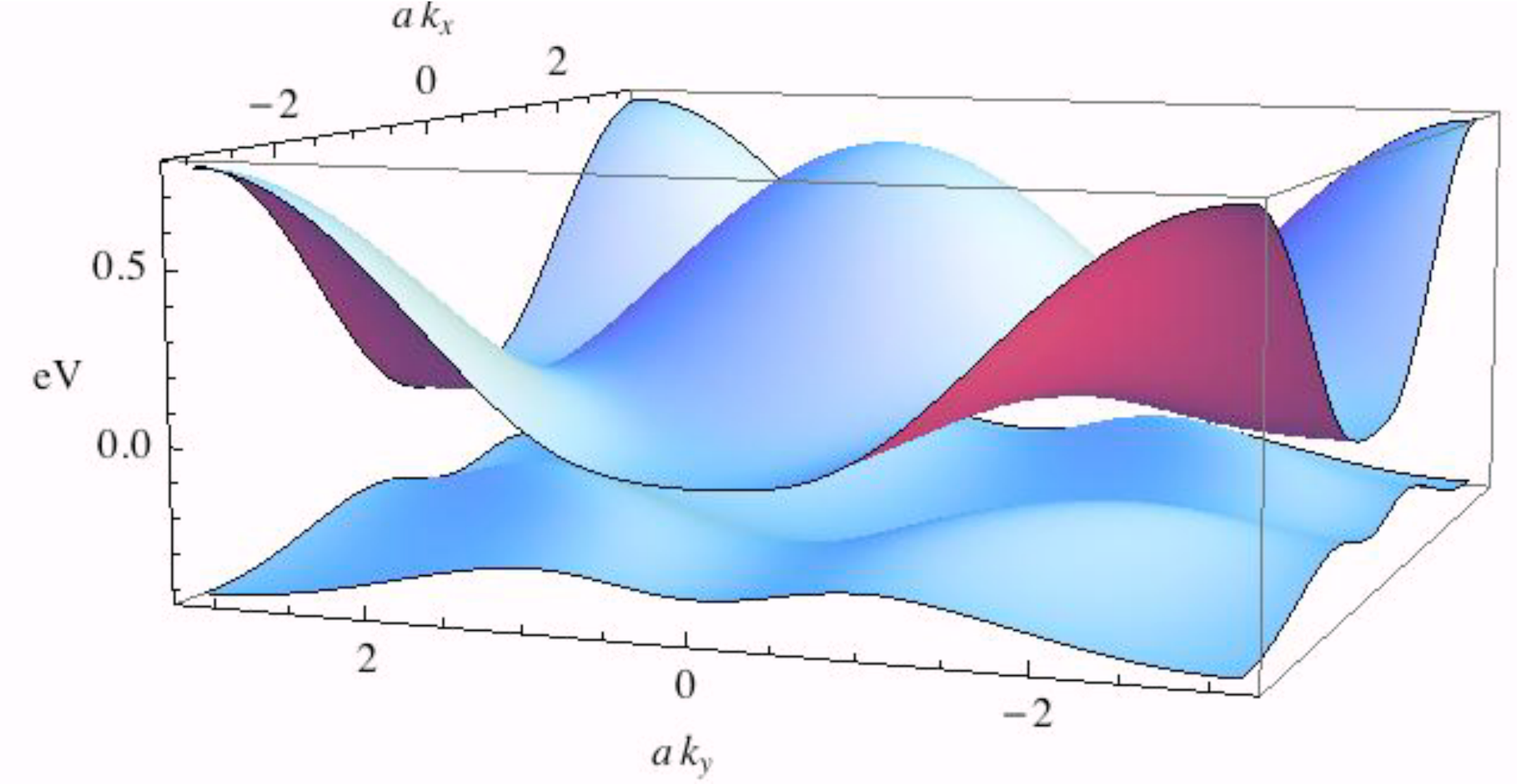}
\caption{Energy spectra, $\lambda_{k,\pm}+\mu$,  corresponding  to st-DDW. Here, for illustration, we have chosen $W_{0}=t$ and $\Delta_{0}=-t$ and for illustration the band parameters are $t=0.15 eV$, $t'=0.3 t$. For the chemical potential, $\mu$, anywhere within the spectral gap, the lower band is exactly a half-filled and the system is a Mott insulator, unlike the semimetallic DDW at half-filling.}
\label{band}
\end{center}
\end{figure}

In the absence of an  external magnetic field, the Hamiltonian is
\begin{equation}
\mathcal{H}_{d\pm i d}-\mu N =\sum_{k} \Psi_{k}^{\dagger} A_k \Psi_k,
\end{equation}
 where the summation is over the reduced Brillouin Zone (RBZ) bounded by $k_y \pm k_x = \pm\pi$, and the spinor, $ \Psi_{k}^{\dagger}$, is defined as $(c_{k,\uparrow}^{\dagger}, c_{k+Q,\uparrow}^{\dagger}, c_{k,\downarrow}^{\dagger}, c_{k+Q,\downarrow}^{\dagger}) $. The chemical potential is subtracted for convenience, $N$ being the number of particles.The matrix $A_k$ is
\begin{equation}
 A_{k} =\left(
  \begin{array}{cccc}
   \epsilon_{k}-\mu  & \Delta_k+iW_k      & 0                 & 0             \\
   \Delta_k-iW_k     & \epsilon_{k+Q}-\mu & 0                 & 0             \\
   0                 & 0                  & \epsilon_{k}-\mu  & \Delta_k-iW_k \\
   0                 & 0                  & \Delta_k+iW_k     & \epsilon_{k+Q}-\mu
  \end{array}
\right),
\end{equation}
with a generic set of band parameters, 
\begin{eqnarray}
\epsilon_k &=&\epsilon_{1k}+\epsilon_{2k}\\
\epsilon_{1k}&=&-2t(\cos k_x + \cos k_y), \; \epsilon_{2k}=4t^{\prime} \cos k_x \cos k_y.
\end{eqnarray}
Here $\epsilon_{2k}$ is possible next nearest neighbor hopping.
Each of the two $2\times2$ blocks can be written in terms of two component spinors, $\psi_{k,\sigma}=(c_{k,\sigma}, c_{k+Q,\sigma})^{T}$, where $\sigma=\pm1\equiv (\uparrow,\downarrow)$; for example, for the up spin block we have
the eigenvalues ($\pm$ refers to the upper and the lower bands respectively)
\begin{equation}
 \lambda_{k,\pm}= \epsilon_{2k}-\mu \pm E_{k}, \;
 E_{k}=\sqrt{\epsilon_{1k}^2 + W_k^2 + \Delta_k^2}.
 \label{eq:gap}
\end{equation}
and are plotted in Fig.~\ref{band}; see Ref.~\refcite{Hsu:2011}.
Since up and down spin components are decoupled, the Chern number for each component can be computed separately. Note that while  $(\epsilon_{2k}-\mu)$ is present in the eigenvalues, it cannot enter the eigenvectors, because the identity operator commutes with the Pauli matrices. After diagonalizing the Hamiltonian, we can obtain the eigenvectors
\begin{equation}
\Phi_{\sigma,\pm}({\bf k}) = (u_{\pm}{\it e}^{i \sigma \theta_k/2}, v_{\pm}{\it e}^{-i \sigma \theta_k/2})^{\mathbf{T}},
\end{equation}
 where ($\Theta(x)$ is the step function)
\begin{eqnarray}
u_{\pm}^2&=&{\frac{1}{2}}(1\pm{\frac{\epsilon_{1k}}{E_k}}), \\
v_{\pm}^2&=&{\frac{1}{2}}(1\mp{\frac{\epsilon_{1k}}{E_k}}), \\
\theta_k &=& \arctan({\frac{W_k}{\Delta_k}}) + \pi \Theta(-\Delta_k).
\end{eqnarray}
The Berry curvature, $\vec{\Omega}_{\sigma,\pm}$ is
\begin{equation}
 \vec{\Omega}_{\sigma,\pm} \equiv i \vec{\bigtriangledown}_k \times \langle \Phi_{\sigma,\pm}^{\dagger}({\bf k})| \vec{\bigtriangledown}_k | \Phi_{\sigma,\pm}({\bf k}) \rangle
\end{equation}
From the eigenstates, the Berry curvature can be written as
\begin{eqnarray}
   \vec{\Omega}_{\sigma,\pm}&=& i \vec{\nabla}_k \times
   [ (u_{\pm}^2-v_{\pm}^2) \vec{\nabla}_k (i\sigma {\frac {\theta_k}{2}})].
\end{eqnarray}
Since $u_{\pm}$, $v_{\pm}$, and $\theta_k$ only depend on $k_x$ and $k_y$, only the z component, $\Omega_{\sigma,\pm}$, is non-zero, which is given by
\begin{eqnarray}
 \Omega_{\sigma,\pm} & = & \mp {\frac {\sigma}{2}} [ {\frac {\partial}{\partial k_x}}({\frac {\epsilon_{1k}}{E_k}})  {\frac {\partial \theta_k}{\partial k_y}} - {\frac {\partial}{\partial k_y}}({\frac {\epsilon_{1k}}{E_k}}) {\frac {\partial \theta_k}{\partial k_x}}] \nonumber \\
&=& \pm \sigma {\frac {1}{2E_k^3}}
 \left| \begin{array}{ccc}
  \Delta_k & W_k & \epsilon_{1k} \vspace{0.1in} \\
  {\frac {\partial{\Delta_k}}{\partial k_x}} & {\frac {\partial{W_k}}{\partial k_x}}&{\frac {\partial{\epsilon_{1k}}}{\partial k_x}} \vspace{0.1in} \\
  {\frac {\partial{\Delta_k}}{\partial k_y}} & {\frac {\partial{W_k}}{\partial k_y}}&{\frac {\partial{\epsilon_{1k}}}{\partial k_y}} \vspace{0.1in}
  \end{array}\right|.
\end{eqnarray}
From the above determinant, we can see that the Berry curvature will be zero if one of $\Delta_k$ and $W_k$ is zero, so we need a mixing of $d_{x^2-y^2}$ and $d_{xy}$ to have a non-trivial topological invariant. 
If we define a unit vector $\hat{n}_{\sigma} \equiv \vec{h}_{\sigma}/|\vec{h}_{\sigma}|$, where 
\begin{equation}
\vec{h}_{\sigma} = (\Delta_k, -\sigma W_k, \epsilon_{1k}), 
\end{equation}
the Berry curvature can be written as
\begin{eqnarray}
 \Omega_{\sigma,\pm} &=& \mp {\frac {1}{2}} \hat{n}_{\sigma} \cdot ( {\frac {\partial{\hat{n}_{\sigma}}}{\partial k_x}} \times {\frac {\partial{\hat{n}_{\sigma}}}{\partial k_y}}).
\end{eqnarray}
More explicitly, the Chern numbers are
\begin{equation}
\begin{split}
N_{\sigma,\pm} &=  \int_{RBZ} {\frac{d^2\it{k}}{2 \pi }} \Omega_{\sigma,\pm}\\
& = \pm \sigma \int_{RBZ} {\frac{d^2\it{k}}{2 \pi }} {\frac {t W_0 \Delta_0} {E_k^3}} (\sin^2k_y + \sin^2k_x \cos^2k_y) \\
 &= \pm \sigma.
 \end{split}
\end{equation}
We can focus on the lower band as long as there is a gap between the upper and the lower bands. Then,
\begin{eqnarray}
  N & = &  N_{\uparrow,-} +  N_{\downarrow,-} = 0 \\
  N_{\text{spin}} & = & N_{\uparrow,-} -  N_{\downarrow,-} = (-1)-1 = -2
\end{eqnarray}
irrespective of the dimensionful parameters. Note, however, that the Chern numbers vanish unless both $\Delta_{0}$ and $W_{0}$ are non-vanishing. The quantization holds for a range of chemical potential $\mu$  within the gap. 

For the fully gapped case, the ratio of the dimensions of the quantized spin Hall conductance to the quantized Hall conductance should be the same as the ratio of the  spin to the charge carried by a particle, since in two dimensions  both quantities have the scale dependence $L^{d-2}$, that is,   
\begin{equation}
  \frac{[\sigma_{xy}^{\text{spin}}]}{[\sigma_{xy}]}  = \frac { {\frac{\hbar}{2}} }{e} .
\end{equation}
So, the quantized spin Hall conductance will be
\begin{eqnarray}
  \sigma_{xy}^{\text{spin}} = -  {\frac {e^2}{h}} {\frac {\hbar}{2e}} \, N_{\text{spin}} = {\frac {e}{2\pi}}    
\end{eqnarray}
As long as time-reversal symmetry is preserved, we will still have Kramers degeneracy in our Hartree-Fock state, and therefore the edge modes resulting from topology  will remain protected. We will dub this state as quantum spin-Hall insulator (QSHI).

 \section{\label{Sec:action}Low Energy Effective Action}

We now go beyond mean field theory.
For simplicity consider  the Hamiltonian (addition of longer ranged hopping will not change our conclusions~\cite{Hsu:2011})
\begin{equation}
\mathcal{H}= \sum_{k,\alpha,\beta} \psi_{k,\alpha}^{\dagger} \left[  \delta_{\alpha\beta} (\tau^{z} \epsilon_{1k} +
 \tau^{x} \Delta_k)
- (\vec{\sigma} \cdot \hat{N})_{\alpha\beta} \tau^{y} W_k \right]\psi_{k\beta},
\end{equation}
where the summation is over the RBZ. Here $\tau^{i}$ ($i=x,y,z$) are Pauli matrices acting on the two-component spinor. We expand around the points $K_1 \equiv(\frac{\pi}{2},\frac{\pi}{2})$ and $K_2 \equiv(-\frac{\pi}{2},\frac{\pi}{2})$,  the two distinct nodal points in the absence of the $d_{xy}$ term, and $K_3 \equiv(0,\pi)$,  the nodal point in the absence of the $d_{x^2-y^2}$ term. 
This allows us to develop an effective \text{low} energy theory by separating the fast modes from the s\text{low} modes. 

After that we make a sequence of transformations for simplicity: (1) transform the Hamiltonian to the real space, which al\text{low}s us to formulate the skyrmion problem; (2) perform a $\pi/2$ rotation along the $\tau^y$-direction, which al\text{low}s us to match to the notation of Ref.~\cite{Nersesyan:1991} for the convenience of the reader; (3) label  $\psi_{K_i+q,\alpha}$ by $\psi_{i\alpha}$, since $K_{i}$ is now a redundant notation; (4) construct the imaginary time effective action, with the definition $\bar{\psi} \equiv -i \psi^{\dagger} \tau ^z$. Finally, after suppressing the spin indices, and with the definitions
$\gamma^0 \equiv \tau^z$, $\gamma^x \equiv \tau^y$, and $\gamma^y \equiv -\tau^x$, we obtain the effective action in a more compact notation: 

\begin{equation}
\begin{split}
S &= \sum_{j=1,2}\int d^3 x\; \bar{\psi}_{j} \bigg[ 
-i \gamma^0 \partial_{\tau}
- 2it  \gamma^x (\eta_{j}\partial_{x}+\partial_{y})\\ 
&+ i\frac{W_0}{2} (\vec{\sigma} \cdot \hat{N}) \gamma^y 
(-\eta_{j}\partial_{x}+\partial_{y})  + i \eta_{j}\Delta_0\bigg] \psi_{j} \nonumber \\
&+\int d^3 x \bar{\psi}_{3} \bigg[ 
-i \gamma^0 \partial_{\tau}
- W_0 (\vec{\sigma} \cdot \hat{N}) \gamma^y \bigg] \psi_{3},
\end{split}
\end{equation}
where $\eta_{1}=1$ and $\eta_{2}=-1$. There is no spatial derivative in the $\psi_{3}$ term since the expansion around the  point $K_{3}=(0,\pi)$ is 
\begin{equation}
W_{K_3+q} = \frac{W_0}{2} (2-\frac{q_x^2}{2}-\frac{q_y^2}{2} +\cdots),
\end{equation}
where the second (and higher) order derivative terms are irrelevant operators. The first term behaves as a mass term at the $K_3$ point. 
\subsection{\label{Sec:charge}The charge and spin of a skyrmion}

First we will compute the charge of the skyrmions in the system.~\cite{Grover:2008} Consider the action around $K_1=(\frac{\pi}{2},\frac{\pi}{2})$ when the order parameter is uniform (say, $\hat{N}=\hat{z}$). The results for $K_2=(-\frac{\pi}{2},\frac{\pi}{2})$  and $K_3=(0,\pi)$
 follow identically. We have shown  above that  in this case the non-trivial topology leads to a quantized spin Hall conductance in  st-DDWstate~\cite{Hsu:2011} as long as the system is fully gapped. The spin quantum Hall effect implies that the external gauge fields $A^c$ and $A^s$ couple to charge and spin currents, respectively. Then,
 \begin{equation}
 \begin{split}
S_1[A^c,A^s] &= \int d^3 x \bar{\psi}_{1} \bigg[ 
-i \gamma^0 \partial_{\tau} + \gamma^0 (A^c_{\tau}+\frac{\sigma^z}{2}A^s_{\tau})
- 2it  \gamma^x (\partial_{x}+\partial_{y})\\
& + 2t \gamma^x (A^c_{x}+\frac{\sigma^z}{2}A^s_{x} + A^c_{y}+\frac{\sigma^z}{2}A^s_{y})  
+ i\frac{W_0}{2}  \sigma^z \gamma^y (-\partial_{x}+\partial_{y})\\  
&- \frac{W_0}{2}  \sigma^z \gamma^y (-A^c_{x}-\frac{\sigma^z}{2}A^s_{x} + A^c_{y}+\frac{\sigma^z}{2}A^s_{y})
+ i \Delta_0   \bigg] \psi_{1}. 
\label{Eq:min}
\end{split}
\end{equation}
The non-vanishing transverse spin conductance implies that the low energy effective action obtained from integrating out the fermions for the gauge fields is given by
\begin{eqnarray}
S_{1,\text{eff}}=\frac{i}{2\pi} \int d^3\it{x} \epsilon^{\mu\nu\lambda}A^c_{\mu}\partial_{\nu}A^s_{\lambda},
\end{eqnarray}
and the charge current is induced by the spin gauge field
\begin{eqnarray}
j^c_{\mu}= \frac{1}{2\pi} \epsilon^{\mu\nu\lambda}\partial_{\nu}A^s_{\lambda}.
\end{eqnarray}

Consider now a static configuration of the $\hat{N}$ field with unit Pontryagin index in the polar coordinate $(r,\theta)$: 
\begin{equation}
\hat{N}(r,\theta)=\left[ \sin \alpha(r) \cos \theta, \sin \alpha(r) \sin \theta, \cos \alpha(r) \right]
\end{equation}
with the boundary conditions $\alpha(r=0)=0$ and $\alpha(r\rightarrow \infty)=\pi$. 
This field configuration associates with a Skyrmion, and now the action is
\begin{equation}
S_1 = \int d^3 x \bar{\psi}_{1}\bigg[ -i \gamma^{0}\partial_{\tau}  -i 2t \gamma^x (\partial_{x}+\partial_{y}) 
+i \frac{W_0}{2} (\vec{\sigma} \cdot \hat{N})\gamma^y (-\partial_{x}+\partial_{y}) +i\Delta_0 \bigg] \psi_{1} \label{Eq:Seff_1}
\end{equation}
We can perform a unitary transformation at all points in space such that
\[
U^{\dagger} (\vec{\sigma} \cdot \hat{N}) U = \sigma^{z},
\]
and define $\psi=U \psi'$, and $\bar{\psi}=\bar{\psi}'U^{\dagger}$. Plugging into the above equation, we obtain
\begin{equation}
\begin{split}
S_1 &=\int d^3 x \bar{\psi}_{1}' \bigg[
-i \gamma^0 \partial_{\tau} - 2it  \gamma^x (\partial_{x}+\partial_{y}) 
+ i\frac{W_0}{2} \sigma^z \gamma^y (-\partial_{x}+\partial_{y})  
+ i \Delta_0 \bigg] \psi'_{1}\\
&+\int d^3 x \bar{\psi}_{1}' \bigg[ -i \gamma^{0}( U^{\dagger}\partial_{\tau} U) -2it \gamma^x (U^{\dagger}\partial_{x}U+U^{\dagger}\partial_{y}U)\\ 
&+i \frac{W_0}{2} \sigma^z \gamma^y (-U^{\dagger}\partial_{x}U+U^{\dagger}\partial_{y}U) \bigg] \psi'_{1}
\label{Eq:rot}
\end{split}
\end{equation}
To proceed, we write down the explicit form for $U(r,\theta)$, which is
\[
U(r,\theta)=\left(
\begin{array}{cc}
\cos \frac{\alpha(r)}{2} & -\sin \frac{\alpha(r)}{2} e^{-i\theta} \\
\sin \frac{\alpha(r)}{2} e^{i\theta} & \cos \frac{\alpha(r)}{2}
\end{array}
\right),
\]
and after applying the boundary conditions as $r\rightarrow \infty$, we have
\begin{eqnarray}
U(r\rightarrow \infty,\theta)=\left(
\begin{array}{cc}
0 & - e^{-i\theta}\\
e^{i\theta} & 0
\end{array}
\right).
\end{eqnarray}
Therefore, in the far field limit, we have
\begin{eqnarray}
U^{\dagger}(r\rightarrow \infty,\theta) \partial_{x} U(r\rightarrow \infty,\theta) \nonumber
&=&\left(
\begin{array}{cc}
0 & e^{-i\theta}\\
-e^{i\theta} & 0
\end{array}
\right)
(\frac{-\sin \theta}{r}\partial_{\theta})
\left(
\begin{array}{cc}
0 & - e^{-i\theta}\\
e^{i\theta} & 0
\end{array}
\right) \nonumber \\
&=&  (\frac{-i\sin \theta}{r})\sigma^z 
\end{eqnarray}
\begin{eqnarray}
U^{\dagger}(r\rightarrow \infty,\theta) \partial_{y} U(r\rightarrow \infty,\theta) \nonumber
&=&\left(
\begin{array}{cc}
0 & e^{-i\theta}\\
-e^{i\theta} & 0
\end{array}
\right)
(\frac{\cos \theta}{r}\partial_{\theta})
\left(
\begin{array}{cc}
0 & - e^{-i\theta}\\
e^{i\theta} & 0
\end{array}
\right) \nonumber \\
&=&  (\frac{i\cos \theta}{r})\sigma^z 
\end{eqnarray}
Putting into Eq.(\ref{Eq:rot}), now we have
\begin{eqnarray}
S_1
&=& \int d^3 x \bar{\psi}_{1}' \left[
-i \gamma^0 \partial_{\tau} - 2it  \gamma^x (\partial_{x}+\partial_{y}) 
+ i\frac{W_0}{2} \sigma^z \gamma^y (-\partial_{x}+\partial_{y})  
+ i \Delta_0 \right] \psi'_{1} \nonumber\\
&& +\int d^3 x \bar{\psi}_{1}' \left[ 2t \gamma^x 
( f_x + f_y  )  + \frac{W_0}{2} \sigma^z \gamma^y ( f_x - f_y ) \right] \psi'_{1}
\end{eqnarray}
where $f_{\mu} \equiv -i U^{\dagger} \partial_{\mu} U$.
Equating the above equation and Eq.~(\ref{Eq:min}), as $r \to \infty$, we obtain the explicit form of the gauge fields in the new basis in the far field limit:
\begin{eqnarray}
A^c_{x}= A^c_{y}=0,  A^s_{x}= - \frac{2\sin \theta}{r}, A^s_{y}=\frac{2\cos \theta}{r}. 
\end{eqnarray}
In other words, the process of tuning the order parameter from $\sigma^z$ to $[\hat{\sigma} \cdot \hat{N}(r,\theta)]$ in the original basis is equivalent  in the new basis to adding an external spin gauge field
\begin{eqnarray}
\vec{A}^s &=& -\frac{2\sin \theta}{r} \hat{x}+\frac{2\cos \theta}{r} \hat{y} \nonumber \\
&=& \frac{2}{r} \hat{\theta}.
\end{eqnarray}

A Skyrmion with unit Pontryagin index in the $i\sigma d_{x^2-y^2}+d_{xy}$ state  induces a spin gauge field $\vec{A}^s=\frac{2}{r} \hat{\theta}$. The total flux of this gauge field is
\begin{eqnarray}
\Phi^s &=&\int d^2x \hat{z} \cdot \vec{\nabla} \times \vec{A}^s \nonumber \\
&=& \int^{2\pi}_{0} r d \theta \cdot (\frac{2}{r}) =4\pi.
\end{eqnarray}

Suppose we adiabatically construct the Skyrmion configuration $\hat{N}(r,\theta)$ from the ground state $\hat{z}$ in a very large time period $\tau_p \rightarrow \infty$. During the process, we effectively thread a spin gauge flux of $4\pi$ adiabatically into the system. The transverse spin Hall conductance implies that a radial current $j^c_{r}$ will be induced by the $4\pi$ spin gauge flux of $\vec{A}^s(t)$, which is now time-dependent: $\vec{A}^s(t=0)=0$ and $\vec{A}^s(t=\tau_p)=\vec{A}^s$. That is,
\begin{eqnarray}
j^c_{r}(t) = -\frac{1}{2\pi} \partial_t A^s_{\theta}(t).
\end{eqnarray}
As a result, charge will be transferred from the center to the boundary, and the total charge transferred during the process can be computed by performing an integral over the boundary and time:
\begin{eqnarray}
Q^c&=&\int^{\tau_p}_{0} dt \int^{2\pi}_{0} r d\theta j^c_r(t) \nonumber \\
&=& - \int^{2\pi}_{0} r d\theta \frac{1}{2\pi} \left[ A^s_{\theta}(\tau_{p})-A^s_{\theta}(0) \right] \nonumber \\
&=& -\frac{2}{2\pi} (2\pi) = -2.
\end{eqnarray}
Therefore, after putting back the unit of charge $e$ to the expression, we obtain a Skyrmion with charge $2e$. The result
is identical for the nodal point $K_{2}$. Since $U^{\dagger} \partial_{\tau} U = 0$, $S_3$ terms do not couple to the gauge fields generated by the Skyrmionic texture. As a result, adding $S_3$ terms does not affect the gauge fields obtained from $S_1$ and $S_2$ terms:
\begin{equation}
\vec{A}^c = 0 ; \; \vec{A}^s = \frac{2}{r} \hat{\theta}.
\end{equation}
Therefore, we have our final result: a Skyrmion in the st-DDW  system carries charge $2e$.

One can also verify the adiabatic result by a different method  by   computing  the Chern numbers.~\cite{Yakovenko:1997} 
The charge and spin of the skyrmions are associated with the coefficients of the Chern-Simons terms by the following relations: $Q_{\text{skyrmion}}= C_{2}e$ and $S_{\text{skyrmion}}=C_{1} \frac{\hbar}{2}$,
where $C_{1}$ and $C_{2}$ are \begin{eqnarray}
C_{1}&=&\frac{\epsilon_{\mu\nu\lambda}}{24\pi^2} {\text{Tr}} \left[ \int d^3k 
G \frac{\partial G^{-1}}{\partial k_{\mu}} 
G \frac{\partial G^{-1}}{\partial k_{\nu}}
G \frac{\partial G^{-1}}{\partial k_{\lambda}} \right], \\
C_{2}&=&\frac{\epsilon_{\mu\nu\lambda}}{24\pi^2}  {\text{Tr}} \left[ \int d^3k (\vec{\sigma} \cdot \hat{z})
G \frac{\partial G^{-1}}{\partial k_{\mu}} 
G \frac{\partial G^{-1}}{\partial k_{\nu}}
G \frac{\partial G^{-1}}{\partial k_{\lambda}} \right],
\end{eqnarray}
where $G$ is the matrix Green's function and the trace is taken over the spin index $\sigma$ and other discrete indices.

If the Green's function matrix is diagonal in the spin index, then the Chern-Simons coefficients for up and down spins can be computed separately. 
\begin{equation}
{\cal N}(G_{\sigma})=\frac{\epsilon_{\mu\nu\lambda}}{24\pi^2} \text{Tr} \left[ \int d^3k 
G_{\sigma} \frac{\partial G_{\sigma}^{-1}}{\partial k_{\mu}} 
G_{\sigma} \frac{\partial G_{\sigma}^{-1}}{\partial k_{\nu}}
G_{\sigma} \frac{\partial G_{\sigma}^{-1}}{\partial k_{\lambda}} \right],
\end{equation}
and
$C_{1}= {\cal N}(G_{\uparrow}) + {\cal N}(G_{\downarrow})$,
$C_{2}= {\cal N}(G_{\uparrow}) - {\cal N}(G_{\downarrow})$.
It can be shown  that 
\begin{eqnarray}
 G_{\sigma}^{-1}= i \omega \hat{I} - \hat{\tau} \cdot \vec{h}_{\sigma}
\end{eqnarray}
with $\vec{h}_{\sigma}$ being the Anderson's pseudospin vector~\cite{Anderson:1958} of the Hamiltonian, where for $i\sigma d_{x^2-y^2}+d_{xy}$ system, we have $\vec{h}_{\sigma}$ defined above. The Chern-Simons coefficient for spin $\sigma$ can be written as
\begin{eqnarray}
{\cal N}(G_{\sigma})= -\int \frac{d^2k}{4\pi} 
\hat{h}_{\sigma} \cdot
\frac{\partial\hat{h}_{\sigma}}{\partial k_{x}} \times 
\frac{\partial\hat{h}_{\sigma}}{\partial k_{y}},  
\end{eqnarray}
where $\hat{h}_{\sigma} \equiv \vec{h}_{\sigma}/|\vec{h}_{\sigma}|$ is a unit vector .  Here  $C_1$ and $C_2$ are the total Chern number $N$ and
the spin Chern number $N_{\textrm{spin}}$ defined in the previous section, respectively. Explicitly, $C_{1}= -1 + 1 = 0$ and $C_{2}= -1 - 1 = -2$;  thus the results are the same as above.

Because a Skyrmion in the system carries integer spin, it obeys bosonic statistics and may undergo Bose-Einstein condensate. As a result, the charge-$2e$ Skyrmion condensate will lead to a superconducting phase transition.  But what about its orbital angular momentum? It can be shown that it is zero~\cite{Hsu:2013} resulting in a $s$-wave singlet state. This is a bit surprising given the original $d$-wave form factor. 
\subsection{The non-linear $\sigma$-model}
It is also possible to derive the corresponding non-linear $\sigma$-model, which also demonstrates the existence of skyrmions,
as we have neglected the hedgehog contribution because skyrmionic excitations have lower energy than the particle-hole excitations required for tunneling between the layers.
Finally we are ready to write down the non-linear $\sigma-$model. The effective action for the field $\hat{N}$ can be written as (correcting a mistake in Ref.\refcite{Hsu:2013})
\begin{align}
S_{\mathrm{eff}} & \approx \frac{1}{g_1} \int d\tau d^2 x [ |\partial_x \hat{N}|^2 +|\partial_y \hat{N}|^2 ] + \frac{1}{g_3} \int d\tau d^2 x |\partial_{\tau} \hat{N}|^2 \\
 & =\frac{1}{g} \int d\tau d^2 x [ |\partial_{\tau} \hat{N}|^2 + v_s^2 \; (|\partial_x \hat{N}|^2 +|\partial_y \hat{N}|^2) ]
\end{align}
with the identification
\begin{align}
\frac{1}{g}& \equiv \frac{1}{g_3} = \sum_{\tilde{k}} \frac{W_0^2}{2(k_0^2 + W_0^2)^2} \\
v_s^2 & \equiv \frac{g_3}{g_1} = \frac{\sum_{\tilde{k}} \frac{W_0^2}{2[k_0^2 + E_{k}^2]}} {\sum_{\tilde{k}} \frac{W_0^2}{2(k_0^2 + W_0^2)^2}}
\end{align}
where $E_k^2=4 t^2[ k_x + k_y]^2 + \frac{W_0^2}{4}[- k_x + k_y]^2 +\Delta_0^2$, and $\tilde{k}=(k_0,k_x,k_y)$. 
We can rescale the spatial coordinates by absorbing the coefficient $v_s^2$ and rewrite the effective action in a more compact form
\begin{gather}
S_{\mathrm{eff}}  \approx \frac{1}{g}\int d\tau d^2 x |\partial_{\mu} \hat{N}|^2
\end{gather}
This completes our derivations of the NL$\sigma$M. 
\section{Chiral $d$-wave Pairing}

We have seen above that Bose condensation of charge $2e$ skyrmions could result in a charged superfluid with zero angular momentum and zero spin. For application to $\mathrm{URu_{2}Si_{2}}$, it is necessary to find a mechanism for chiral $d$-wave superconductor.~\cite{Kasahara:2007,Yano:2008} This will be accomplished by fractionalization of skyrmions into merons and anti-merons. A meron is half a skyrmion. A mapping on the surface of a sphere of the two-dimensional plane  covers only half the solid angle $2\pi$ instead of $4\pi$.

\begin{figure}[htbp]
	\centering
	\includegraphics[width=0.5\linewidth]{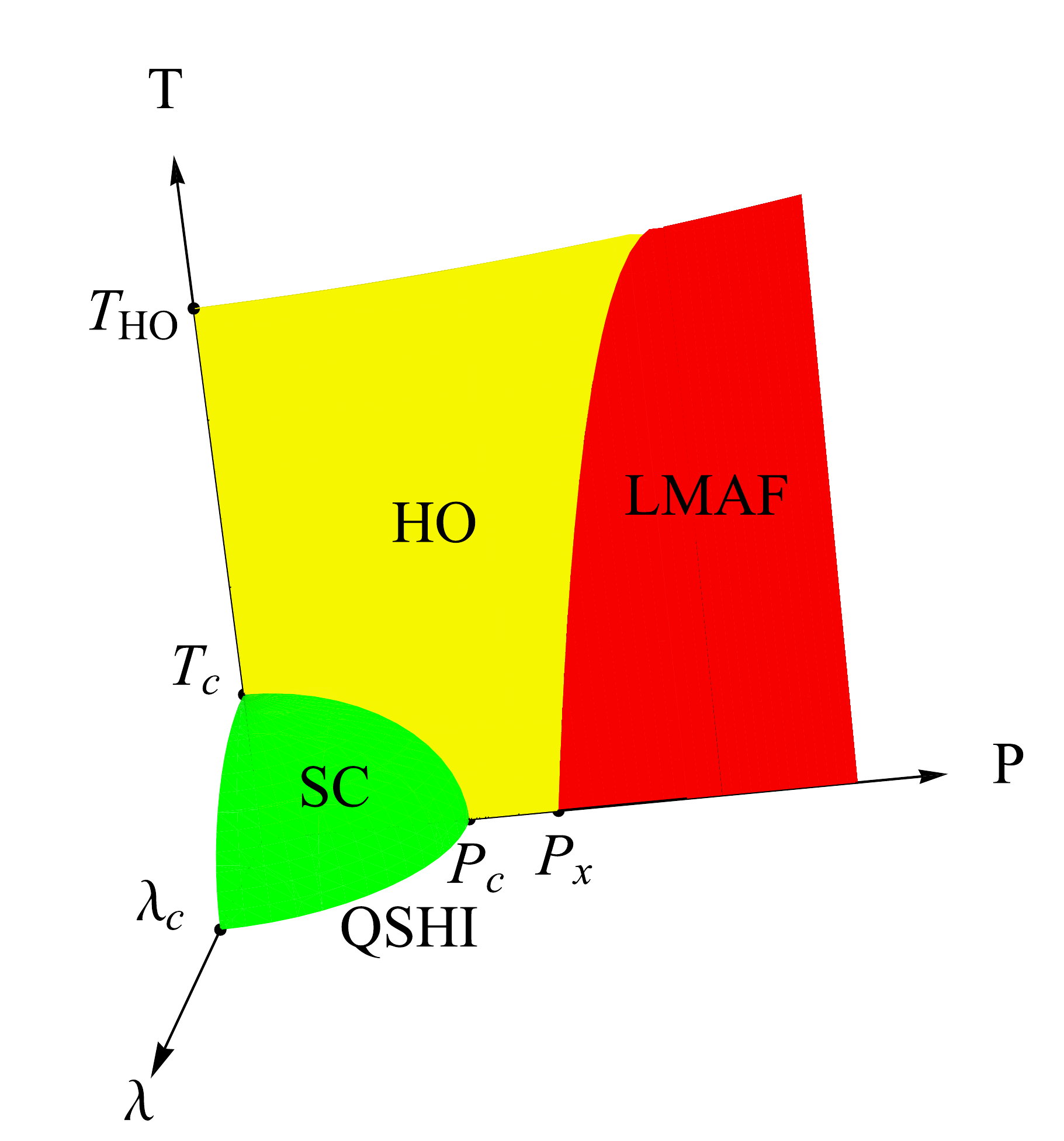} 
	\caption{The proposed phase diagram~\cite{Hsu:2014} with the quantum parameter ($\lambda$), pressure ($P$), and temperature ($T$) axes. Here $\lambda$ is a tuning parameter such that  $W_{0}(\lambda<\lambda_{c})=0$ and $W_{0}(\lambda>\lambda_{c})\neq 0$. $\lambda_{c}$ is a deconfined quantum critical point between the QSHI and superconductivity as $T=P=0$. $T_{\textrm{HO}}$ and $T_{c}$ are the HO and superconducting transition temperatures as $P=\lambda=0$, respectively. Along the $P$ axis, $P_{c}$ indicates the phase transition between the HO and superconducting states, while $P_{x}$ indicates the phase transition between the HO and the large moment antiferromagnetic  states.~\cite{Mydosh:2011,Mydosh:2014} In some literatures $P_{c}$ coincides with $P_{x}$, which, however,  does not affect our main conclusion. In addition, there should be phase boundaries in the $\lambda$-$P$ and $\lambda$-$T$ planes, which are not the main purpose of this work.}
	\label{fig:PD}
\end{figure}

For the purpose of orientation, consider a phase diagram in which we introduce a quantum parameter $\lambda$ in addition to the  parameters pressure, $P$, and temperature $T$, as shown in Fig.~\ref{fig:PD}; see, Ref.~\refcite{Hsu:2014}.
$\lambda$ controls  $W_{0}(\lambda)$ such that $W_{0}(\lambda<\lambda_{c})=0$ and $W_{0}(\lambda>\lambda_{c})\neq 0$. Since an isolated meron costs logarithmically infinite energy for $\lambda > \lambda_{c}$,  merons and antimerons appear as bound pairs in skyrmions. The length scale of the confinement potential grows when approaching the deconfined quantum critical point $\lambda_{c}$, where it diverges. Therefore, the  skyrmions fractionalize into merons and antimerons, because there is no confinement at that point. It is assumed that the hedgehog configurations are  suppressed because the particle-hole excitations are of much higher energy.~\cite{Grover:2008} Therefore the skyrmion number is conserved in the two-dimensional $XY$-planes. 

The state at $T=0,P=0$ is connected, as is the entire superconducting state, by continuity. $\lambda_{c}$ is a deconfined quantum critical point.
The suppression  of hedgehog configurations is crucial to the existence of the deconfined quantum critical point.
These fractional particles, merons and antimerons, are present  at $\lambda_{c}$, but  not  in either side of it. We presume that $\lambda_{c}$ can be computed from a suitable microscopic Hamiltonian; for instance, $\lambda$ may be a function of the on-site Coulomb interaction $U$, the nearest neighbor direct interaction $V$, and the exchange interaction $J$ in an extended Hubbard model.~\cite{Ikeda:1998}

Since merons have topological charge they should have zero overlap with band fermions and therefore cannot be expressed in terms of local band fermonic operators. This is not unprecedented: the creation/annihilation of Laughlin quasiparticles in the fractional quantum Hall effect cannot be expressed as any local function of the band fermions.

\subsection{The interaction term}

\begin{figure}[htbp]
	\centering
	\includegraphics[width=0.5\linewidth]{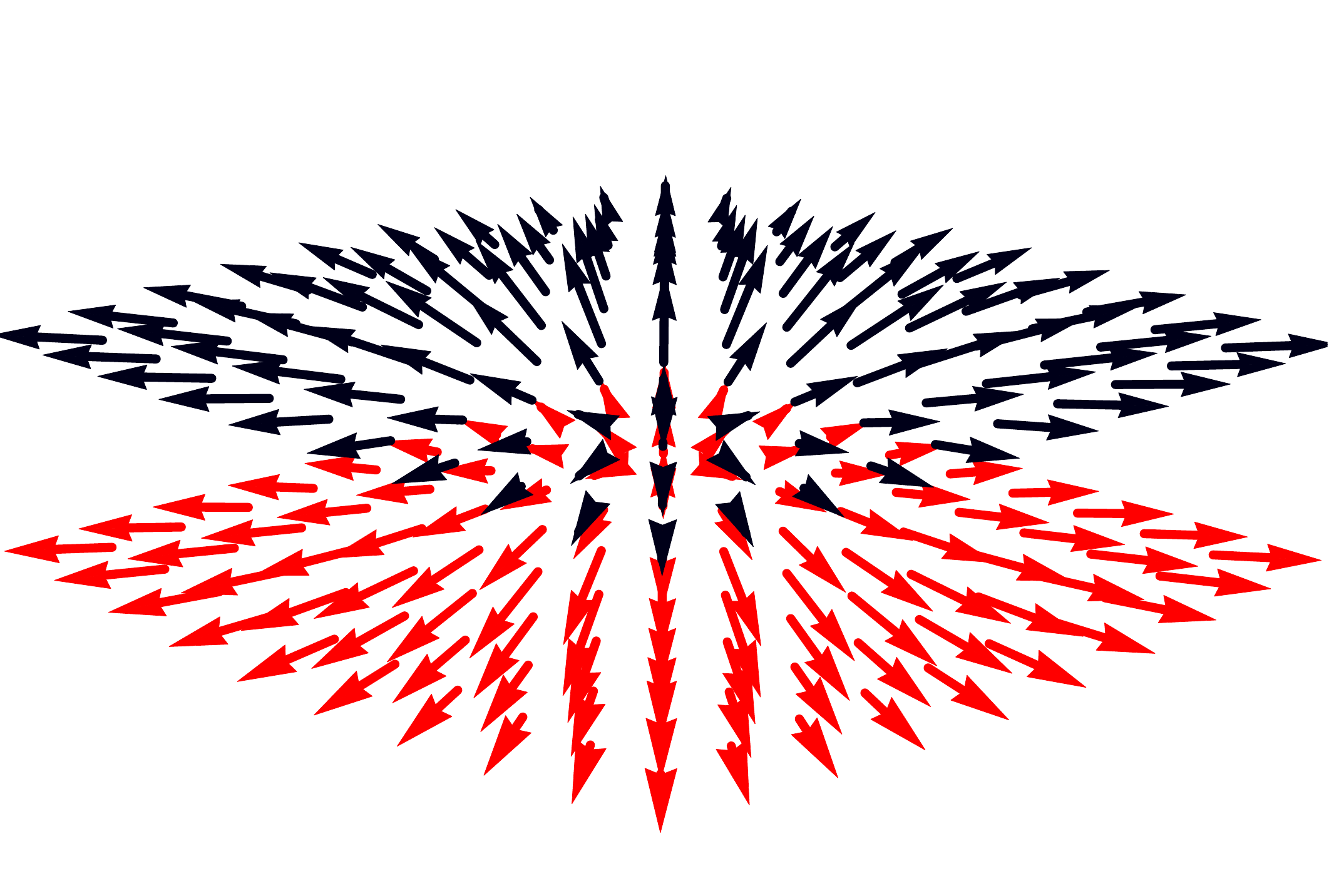} 
	\caption{The merons $\psi_{+,\sigma}^{\dagger}(\vec{r})$ and $\psi_{-,\sigma}^{\dagger}(\vec{r})$. $\psi_{+,\sigma}^{\dagger}(\vec{r})$ creates a meron with $\hat{N}(r\rightarrow 0)=\hat{z}$ and $\hat{N}(r\rightarrow \infty)=\frac{(x,y,0)}{r}$; $\psi_{-,\sigma}^{\dagger}(\vec{r})$ creates a meron with $\hat{N}(r\rightarrow 0)=-\hat{z}$ and $\hat{N}(r\rightarrow \infty)=\frac{(x,y,0)}{r}$. Each meron above is half the skyrmion: A composite of $\psi_{+,\sigma}^{\dagger}(\vec{r})$ and $\psi_{-,\sigma'}(\vec{r})$ makes one skyrmion; see Ref.~\citenum{Hsu:2014}.}
	\label{Fig:composite}
\end{figure}

\begin{figure}[htbp]
	\centering
	\includegraphics[width=0.25\linewidth]{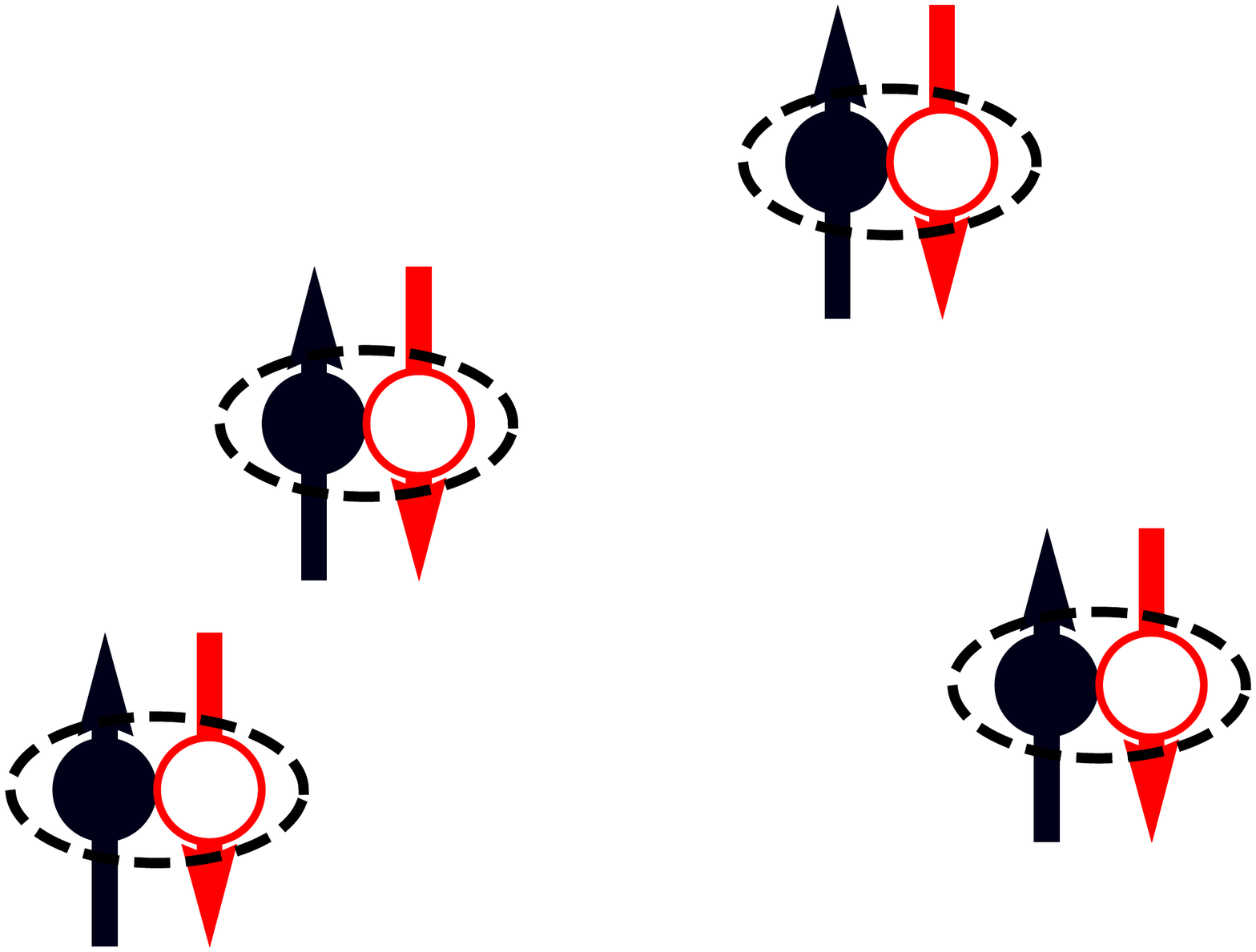}\hspace{0.1\linewidth} 
	\includegraphics[width=0.25\linewidth]{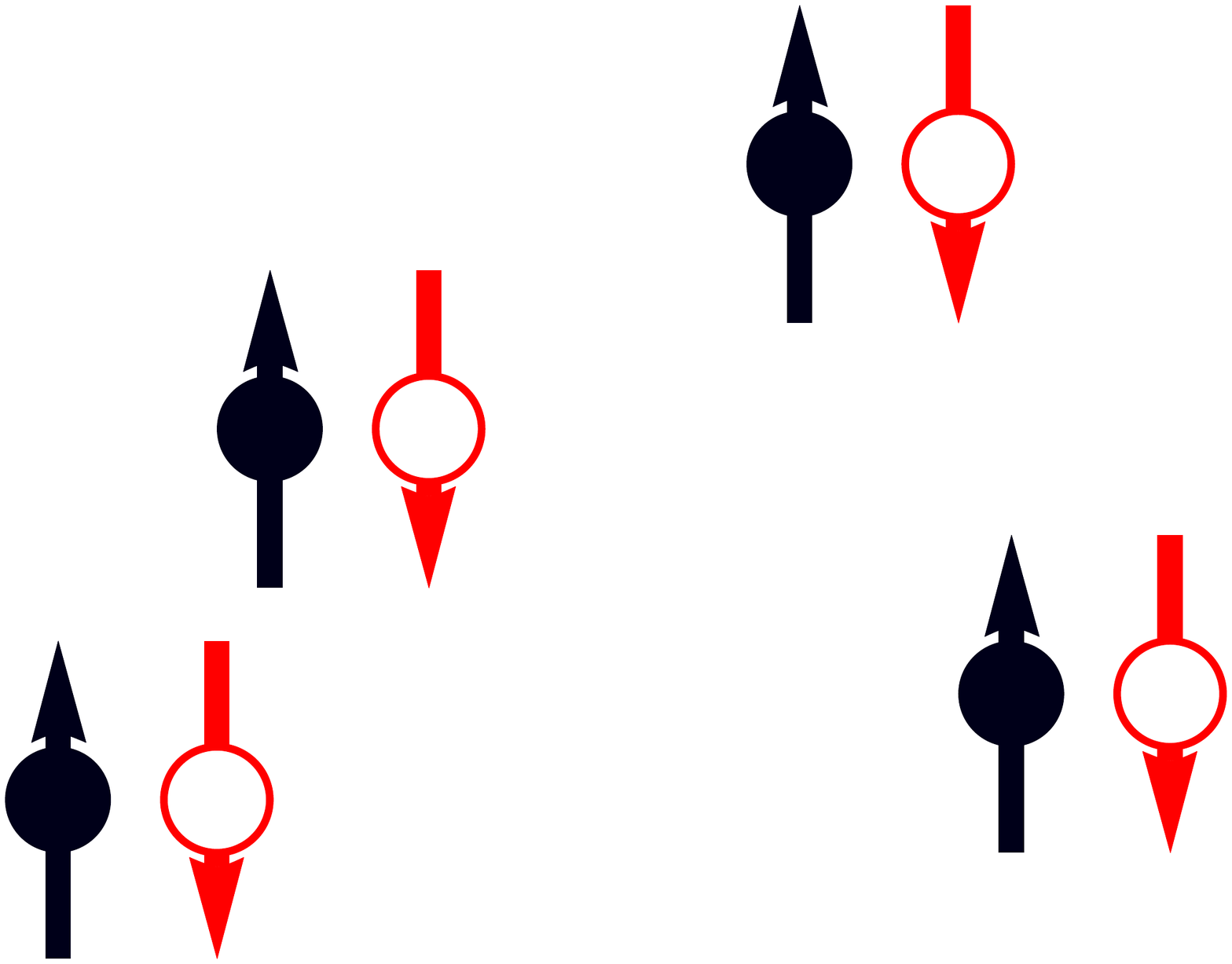}\hspace{0.1\linewidth} 
	\includegraphics[width=0.25\linewidth]{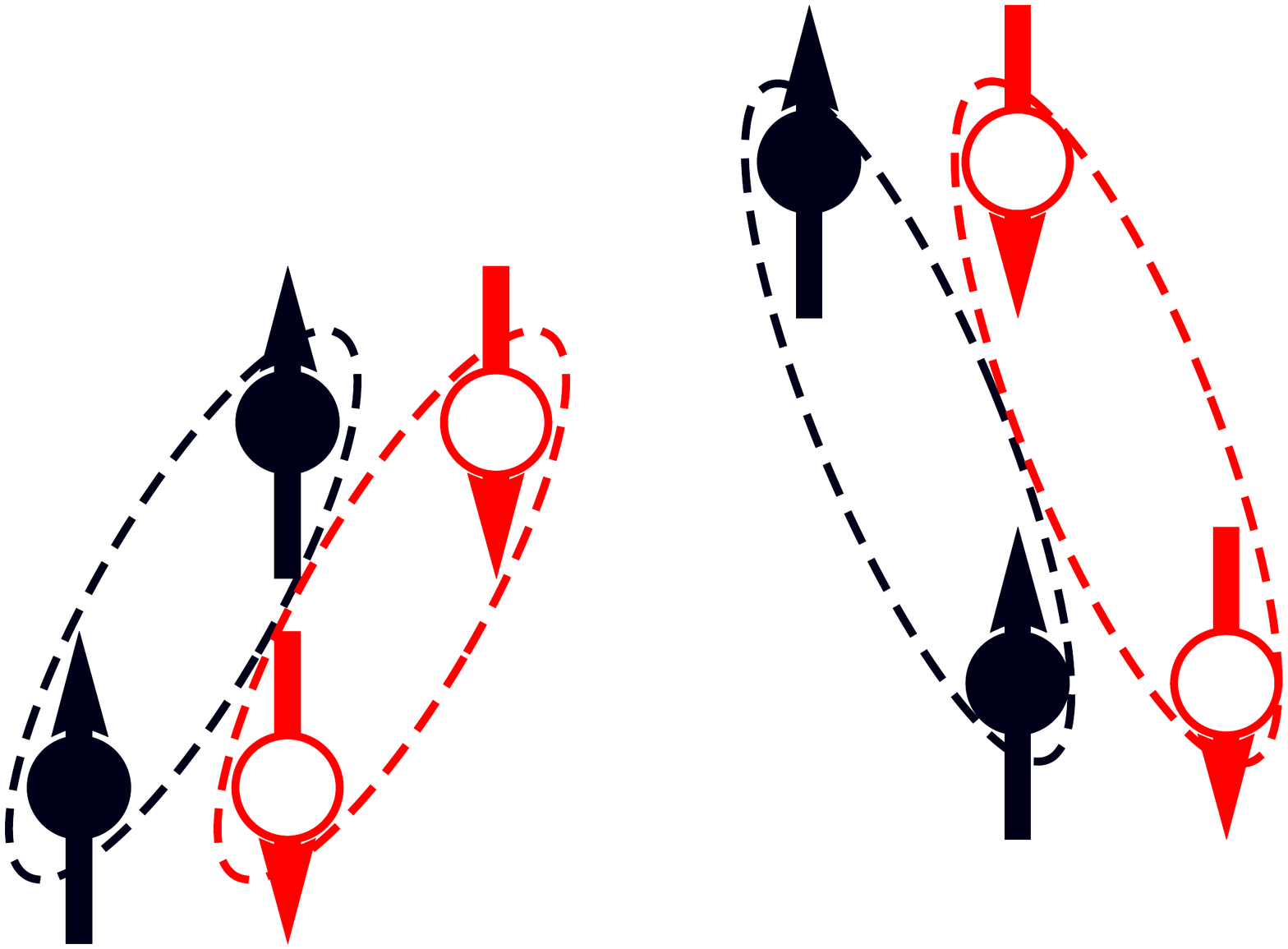}
	\caption{The deconfinement and pairing of the merons.~\cite{Hsu:2014} 
The up (down) arrows indicate a flux of $2\pi(-2\pi)$. The solid (open) circles indicate merons (anti-merons). The colors are associated with the meron texture in Fig.~\ref{Fig:composite}.
Left: in the st-DDW phase, a pair of a meron and an anti-meron is confined in a skyrmion. Middle: at the deconfined quantum critical point, the merons, which are fractions of a skyrmion, emerge. Right: In the superconducting phase, the merons are bound again into a Cooper pair.}
	\label{Fig:deconfine}
\end{figure}

Consider a skyrmion with a flux of $4\pi$ (in unit of $\hbar c/e$), as a composite of a meron with a flux of $2\pi$ and an anti-meron with a flux of $-2\pi$, as shown in Fig.~\ref{Fig:composite}. 
We can imagine that at the critical point, the merons get deconfined. These fractional particles emerge as natural degrees of freedom right at the deconfined quantum critical point~\cite{Senthil:2004}. Thus, we can study the pairing instability which results from the interaction between these fractional particles. These deconfined merons only emerge $\lambda_{c}$, but not on either side of it. In the st-DDW phase, the fractional particles are confined in skyrmions while in the superconducting phase they are bound into a Cooper pairs (Fig.~\ref{Fig:deconfine}; see Ref.~\refcite{Hsu:2014}).

Let $\psi_{s,\sigma}^{\dagger}(\vec{r})$ be the creation operator of a meron at $\vec{r}$, where $s=\pm$ labels the flux of $\pm 2\pi$ and $\sigma=\uparrow$($\downarrow$) for up- (down-) spin. Here $\psi_{\pm,\sigma}^{\dagger}(\vec{r})$ carries charge of $\pm e$, so a skyrmion, a composite of $\psi_{+,\sigma}^{\dagger}(\vec{r})$ and $\psi_{-,\sigma'}(\vec{r})$, carries charge $e-(-e)=2e$ and flux $2\pi-(-2\pi)=4\pi$. The subtraction is because destroying a particle with charge $-e$ and flux $-2\pi$ is equivalent to creating a particle with charge $+e$ and flux $+2\pi$. 
Therefore, a pairing of $\langle \psi_{s,\sigma}^{\dagger}(\vec{r}) \psi_{s',\sigma'}^{\dagger}(\vec{r})\rangle$ results in a charge $2e$ superconductivity for $s=s'$. For singlet pairing of the merons, $\sigma=-\sigma'$. 

Notice that the meron-anti-meron pair which constitute a skyrmion is not the same as the meron-meron pair which lead to the superconductivity. At the deconfined quantum critical point, the merons are not bound in skyrmions, so they can interact with the merons within other skyrmions. This may be the reason why a skymion has zero angular momentum~\cite{Hsu:2013} while a Cooper pair formed by the merons may have nonzero angular momentum. In other words, we are not studying the internal structure of a skyrmion. Instead, we are studying the pairing mechanism due to these fractional particles emerging at the deconfined quantum critical point.

To formulate the interaction between the merons at the critical point, we imagine a situation  where the gauge flux density $\vec{\triangledown} \times \vec{A}$ is produced by a meron. Suppose another meron passes the region of the created  gauge flux, it will feel the Lorentz force. Because these particles carry charges, the motion of the merons also produces charge currents, and a magnetic interaction is  induced between the two particles~\cite{Greiter:1992,Morinari:2006}. We also assume that the length scale of a meron is much smaller than the distance between merons, so that we can treat them as point particles. This is a critical assumption, as is the assumption that in the continuum limit discussed below, the merons or antimerons have kinetic energy $k^{2}/2m$, with a finite mass. Going beyond these assumptions is outside the scope of this paper.

The coupling between the gauge field and the charge current can be described by the imaginary time Lagrangian density, which is
\begin{eqnarray}
{\cal L}_{int} = - i \; j_{\mu} A_{\mu}
\end{eqnarray}
Therefore, the interaction Hamiltonian is
\begin{eqnarray}
{\cal H}_{int} &=&  \int d^2 r \vec{j}(\vec{r}) \cdot \vec{A}(\vec{r}) \nonumber \\
&=&    \sum_{\vec{q}} \vec{j}_{-q} \cdot \vec{A}_q
\label{Eq:interaction}
\end{eqnarray}
The gauge flux is induced by the spin texture, and adding a meron $\psi_{\pm,\sigma}^{\dagger}(\vec{r})$ is equivalent to threading a gauge flux of $\pm 2\pi$ at $\vec{r}$. 

So, the total gauge flux density and the density of the merons are related through the following relation,  
\begin{equation}
\left( \vec{\nabla} \times \vec{A}(\vec{r}) \right)_z =  2\pi  \sum_{s,\sigma} s \psi_{s,\sigma}^{\dagger}(\vec{r}) \psi_{s,\sigma}(\vec{r})
\label{Eq:density},
\end{equation}
where $(\vec{\nabla} \times \vec{A})_z \equiv \partial_x A_y - \partial_y A_x$, and the charge density of the merons is 
\begin{equation}
\rho(\vec{r}) = \sum_{s,\sigma} s\; \psi_{s,\sigma}^{\dagger}(\vec{r}) \psi_{s,\sigma}(\vec{r})
\label{Eq:density2}
\end{equation}
In addition, we have the continuity equation, which describes the relation between the density and the current of the merons,
\begin{equation}
\frac{\partial \rho(\vec{r})}{\partial t} + \vec{\triangledown} \cdot \vec{j}(\vec{r}) = 0
\end{equation}
With the continuity equation, and assuming that the kinetic energy of merons is 
$\epsilon_k = \frac{k^2}{2m}$ with the effective mass of merons $m$,
we may write the current in terms of the $\psi_{\pm,\sigma}^{\dagger}(\vec{r})$ operator,
\begin{equation}
\vec{j}(\vec{r}) = \frac{1}{2mi}  \sum_{s,\sigma} s \left\{
\psi_{s,\sigma}^{\dagger}(\vec{r}) \left[ \vec{\triangledown} \psi_{s,\sigma}(\vec{r}) 
\right]
- \left[ \vec{\triangledown} \psi_{s,\sigma}^{\dagger}(\vec{r}) \right] \psi_{s,\sigma}(\vec{r}) 
\right\}
\label{Eq:current}
\end{equation}
The Fourier transform of Eq.(\ref{Eq:density}) and Eq.(\ref{Eq:current}) gives 
\begin{equation}
\left( i\vec{q} \times \vec{A}_q \right)_z  = 2\pi  \sum_{\vec{k},s,\sigma} s \psi_{k,s,\sigma}^{\dagger} \psi_{k+q,s,\sigma}
\label{Eq:density-FT}
\end{equation} 
and 
\begin{equation}
\vec{j}_q = \frac{1}{m} \sum_{\vec{k},s,\sigma} s (\vec{k} + \frac{\vec{q}}{2} )
\psi_{k,s,\sigma}^{\dagger}  \psi_{k+q,s,\sigma}
\label{Eq:current-FT}.
\end{equation} 
Now the current is expressed in terms of $\psi_{k,s,\sigma}^{\dagger}$. The next step is to write the gauge field in terms of $\psi_{k,s,\sigma}^{\dagger}$ as well. This can be done by choosing the Coulomb gauge, $\vec{\triangledown} \cdot \vec{A}(\vec{r}) = 0$, or $i \vec{q} \cdot \vec{A}_q = 0$. This means we can write $\vec{A}_q$ in the following form,
\begin{equation}
\vec{A}_q = \frac{i}{q^2}A_q \left( q_y \hat{x} - q_x \hat{y}  \right),
\end{equation} 
where $q^2=q_x^2+q_y^2$ and $A_q = |\vec{A}_q| = \sqrt{ A_{q,x}^2+A_{q,y}^2 }$.

Plugging into Eq.(\ref{Eq:density-FT}), we obtain the expression for $A_q$ and hence $\vec{A}_q$,
\begin{equation}
\vec{A}_q = \frac{2\pi i}{q^2}   \sum_{\vec{k},s,\sigma} s \psi_{k,s,\sigma}^{\dagger}  \psi_{k+q,s,\sigma} \left( q_y \hat{x} - q_x \hat{y}  \right).
\label{Eq:gauge-field}
\end{equation}
From Eq.(\ref{Eq:current-FT}), we have  the interaction Hamiltonian 
\begin{equation}
{\cal H}_{int} 
= \frac{2\pi i}{m} \sum_{\vec{q},\vec{k}_1,\vec{k}_2} \sum_{s_1 s_2} \sum_{\sigma_1,\sigma_2}
s_1 s_2 \frac{(\vec{k}_1 \times \vec{q})_z}{q^2} 
\psi_{k_1+q,s_1,\sigma_1}^{\dagger}  \psi_{k_1,s_1,\sigma_1}  \psi_{k_2,s_2,\sigma_2}^{\dagger}  \psi_{k_2+q,s_2,\sigma_2}
\end{equation}
Since we are interested in the Cooper pairing, we may set $\vec{k}_1+\vec{k}_2+\vec{q}=0$ and sum over $\vec{q}$ to obtain 
\begin{equation}
{\cal H}_{int} 
= \frac{2\pi i }{m}  \sum_{\vec{k}_1,\vec{k}_2} \sum_{s_1 s_2} \sum_{\sigma_1,\sigma_2}
s_1 s_2 \frac{-(\vec{k}_1 \times \vec{k}_2)_z}{|\vec{k}_1 + \vec{k}_2|^2} 
\psi_{-k_2,s_1,\sigma_1}^{\dagger}  \psi_{k_1,s_1,\sigma_1}  \psi_{k_2,s_2,\sigma_2}^{\dagger}  \psi_{-k_1,s_2,\sigma_2}
\end{equation}
Setting $\vec{k}_1 = \vec{k}$ and $\vec{k}_2 = -\vec{k}'$, $s_1=s_2=s$, $\sigma_1=-\sigma_2=\sigma$,  for singlet charge $2e$ pairing, the interaction Hamiltonian can be written as
\begin{eqnarray}
{\cal H}_{int} 
&=& \frac{2\pi i}{m} \sum_{\vec{k},\vec{k}'} \sum_{s,\sigma} s^2
\frac{(\vec{k} \times \vec{k}')_z}{|\vec{k} - \vec{k}'|^2} 
\psi_{k',s,\sigma}^{\dagger} \psi_{-k',s,-\sigma}^{\dagger}  \psi_{-k,s,-\sigma} \psi_{k,s,\sigma}
\end{eqnarray}
Under time reversal, we have
\begin{eqnarray}
{\cal H}'_{int} 
&=& -\frac{2\pi i}{m}  \sum_{\vec{k},\vec{k}'} \sum_{s,\sigma}
\frac{(\vec{k} \times \vec{k}')_z}{|\vec{k} - \vec{k}'|^2} 
\psi_{k',s,\sigma}^{\dagger} \psi_{-k',s,-\sigma}^{\dagger}  \psi_{-k,s,-\sigma} \psi_{k,s,\sigma}, \nonumber \\
&\neq& {\cal H}_{int} 
\end{eqnarray}
so the time reversal symmetry is broken in the interaction term, and we expect the superconducting order to break time reversal symmetry.

\subsection{The BCS gap equation}

With the kinetic energy term, the total Hamiltonian is
\begin{eqnarray}
{\cal H}_{total} 
&=& \sum_{\vec{k},s,\sigma} \xi_k  \psi_{k,s,\sigma}^{\dagger}  \psi_{k,s,\sigma}  
+ \sum_{\vec{k},\vec{k}',s} V_{k'k}
\psi_{k',s,\uparrow}^{\dagger} \psi_{-k',s,\downarrow}^{\dagger}  \psi_{-k,s,\downarrow} \psi_{k,s,\uparrow},
\end{eqnarray}
where $\xi_k \equiv \epsilon_k -\mu$ with the chemical potential $\mu$, and the summation over $\sigma$ was carried out. The BCS interaction is given by
\begin{equation}
V_{k'k} \equiv \frac{4\pi i}{m} \frac{(\vec{k} \times \vec{k}')_z}{|\vec{k} - \vec{k}'|^2}
\end{equation}
Here we can see that the $s=+$ and $s=-$ parts are identical, so we will suppress the flux index $s$ in what follows. In other words, we have
\begin{eqnarray}
{\cal H}_{total} 
&=& \sum_{\vec{k},\sigma} \xi_k  \psi_{k,\sigma}^{\dagger}  \psi_{k,\sigma}  
+ \sum_{\vec{k},\vec{k}'} V_{k'k}
\psi_{k',\uparrow}^{\dagger} \psi_{-k',\downarrow}^{\dagger}  \psi_{-k,\downarrow} \psi_{k,\uparrow}
\end{eqnarray}
With  $b_k =  \langle \psi_{-k,\downarrow} \psi_{k,\uparrow} \rangle$ and $b_k^{*} = \langle \psi_{k,\uparrow}^{\dagger} \psi_{-k,\downarrow}^{\dagger} \rangle$,
and defining, as usual,  $\Delta_k^{sc} = - \sum_{k'} V_{kk'} b_{k'}$, the superconducting gap, we get (note the distinction with the amplitude of the $d_{xy}$ gap $\Delta_{k}$ in st-DDW)
\begin{eqnarray}
b_{k'} &\equiv& \langle \psi_{-k',\downarrow} \psi_{k',\uparrow} \rangle \nonumber\\
&=& \frac{\Delta_{k'}^{sc}}{2\sqrt{ \xi_{k'}^2 + |\Delta_{k'}^{sc}|^2}} \left[ 1 - 2 f(\sqrt{ \xi_{k'}^2 + |\Delta_{k'}^{sc}|^2})\right],
\end{eqnarray}
where $f(x)$ is the Fermi-Dirac distribution function.
The BCS gap equation at $T=0$ is therefore
\begin{eqnarray}
\Delta_k^{sc} = - \sum_{k'} V_{kk'} 
\frac{\Delta_{k'}^{sc}}{2\sqrt{\xi_{k'}^2 + |\Delta_{k'}^{sc}|^2 }},
\end{eqnarray}
where $V_{kk'}=-V_{k'k}$. Below we shall also solve the gap equation at finite temperatures. 

Because of the imaginary prefactor of the potential, the $s$ wave pairing would not be a solution since $\Delta_k^{sc}$ cannot be real. The interaction is similar to the one discussed in the half-filled Landau level problem~\cite{Greiter:1992} as well as the one in the context of the hole-doped cuprates~\cite{Morinari:2006}. However, there are some differences we would like to stress. In Ref.~\refcite{Greiter:1992}, the flux attached to a particle is $\pi \epsilon$ instead of $2\pi$ to study fractional statistics  by varying $\epsilon$. The particles in their system are spinless fermions, so they obtained a pairing state with odd-parity. The interaction is also different from the one in Ref.~\refcite{Morinari:2006} because we express $\rho(\vec{r})$ and $\vec{A}(\vec{r})$ in terms of the $\psi_{s,\sigma}(\vec{r})$ operator differently. In Ref.~\refcite{Morinari:2006}, the resulting interaction depends on the sign of $s$, so the $s=\pm$ part leads to a $(d_{x^2-y^2} \mp id_{xy})$ superconductivity, respectively. As a result, the addition of these two components gives a $d_{x^2-y^2}$ superconductivity, but not a chiral state in cuprates. The merons and antimerons provide two identical copies of chiral superconductors.

\subsection{The solution of the  gap equation}

We begin with the ansatz for $l$ wave pairing,
\begin{equation}
\Delta_k^{sc} = |\Delta_k^{sc}| e^{i l \phi_k},
\end{equation}
where $\phi_k$ denotes the direction of the wave vector, and we will choose it to be the angle between $\vec{k}$ and $\vec{k}'$ for simplicity.
Plugging the ansatz into the gap equation, we have
\begin{eqnarray}
|\Delta_k^{sc}| 
&=&  -  \frac{ i}{4\pi m} \int_0^{\infty} k' dk' 
\frac{|\Delta_{k'}^{sc}|}{\sqrt{\xi_{k'}^2 + |\Delta_{k'}^{sc}|^2 }} 
\int_0^{2\pi} d\phi
\frac{\sin \phi e^{il\phi}}{ \lambda_{kk'} -\cos \phi},
\end{eqnarray}
where $\lambda_{kk'} \equiv \frac{k^2+k'^2}{2kk'}$.

The angular integral can be computed  by performing a contour integral in the complex plane. To do this, we set $z=e^{i\phi}$ and get
\begin{eqnarray}
I_l(\lambda) 
&=& -\oint \frac{dz}{z} 
\frac{ (z^2-1) z^{l}}{ 2\lambda z - z^2 -1 },
\end{eqnarray}
where we have used $d \phi = \frac{dz}{iz}$and $\lambda \equiv \lambda_{kk'}$ for simplicity, and converted the $\phi$ integral into a contour integral around the origin with unit radius
\begin{eqnarray}
I_l(\lambda) 
&=& \oint dz 
\frac{ (z^2-1) z^{l-1}}{ (z-\lambda-\sqrt{\lambda^2-1})(z-\lambda+\sqrt{\lambda^2-1}) },
\end{eqnarray}
where the poles are at $z=\lambda \pm \sqrt{\lambda^2-1}$. Since $z=\lambda + \sqrt{\lambda^2-1}\ge 1$ for $\lambda \ge 1$, it is not enclosed by the contour. Thus, only $z=\lambda - \sqrt{\lambda^2-1}$ contributes to the integral, and we get
\begin{eqnarray}
I_l(\lambda) 
&=& 2\pi i 
(\lambda - \sqrt{\lambda^2-1})^{l}
\end{eqnarray}

As $\lambda\equiv\lambda_{kk'} = \frac{k^2+k'^2}{2kk'}$, we have
\begin{eqnarray}
\lambda_{kk'} - \sqrt{\lambda_{kk'}^2-1} 
&=& \left\{ 
\begin{array}{ll}
\left(\frac{k'}{k}\right) & \textrm{ as } k>k' \\
\left(\frac{k}{k'}\right) & \textrm{ as } k<k'
\end{array}
\right.
\end{eqnarray}
So, 
\begin{eqnarray}
I_l (\lambda_{kk'}) &=& 2\pi i \times \left\{ 
\begin{array}{ll}
\left(\frac{k'}{k}\right)^{l} & \textrm{ as } k>k' \\
\left(\frac{k}{k'}\right)^{l} & \textrm{ as } k<k'
\end{array}
\right.
\end{eqnarray}
Putting back into the gap equation, we have
\begin{eqnarray}
|\Delta_k^{sc}| 
&=& \frac{ 1 }{2m} \left[
\int_0^{k}  dk' \left(\frac{k'}{k}\right)^{l}
\frac{k'|\Delta_{k'}^{sc}|}{\sqrt{\xi_{k'}^2 + |\Delta_{k'}^{sc}|^2 }}  
+\int_k^{\infty}  dk' \left(\frac{k}{k'}\right)^{l}
\frac{k'|\Delta_{k'}^{sc}|}{\sqrt{\xi_{k'}^2 + |\Delta_{k'}^{sc}|^2 }}  
\right]
\label{Eq:gap1}
\end{eqnarray}
Here we can see that $|\Delta_k^{sc}| \propto k^{-l}$ as $k \rightarrow \infty$ and $|\Delta_k^{sc}| \propto k^{l}$ as $k \rightarrow 0$, so we take the ansatz
\begin{eqnarray}
|\Delta_k^{sc}| &=& \left\{ 
\begin{array}{ll}
\Delta_{F}^{sc} \left(\frac{k}{k_{F}}\right)^{l} & \textrm{ as } k \le k_{F}, \\
\Delta_{F}^{sc} \left(\frac{k_{F}}{k}\right)^{l} & \textrm{ as } k \ge k_{F},
\end{array}
\right.
\label{Eq:gap-ansatz}
\end{eqnarray}
where $\Delta_{F}^{sc} \equiv |\Delta_{k_{F}}^{sc}|$ with the Fermi wave vector $k_{F}$.  
Assuming $k=k_{F}$ and plugging Eq.(\ref{Eq:gap-ansatz}) into Eq.(\ref{Eq:gap1}), we obtain
\begin{equation}
\Delta_F^{sc} 
= \frac{ \Delta_F^{sc} }{2m} \bigg[
\int_0^{k_F}  dk' 
\frac{k' \left( \frac{k'}{k_F} \right)^{2l} }{\sqrt{(\frac{k'^2}{2m}-\frac{k_F^2}{2m})^2 + (\Delta_F^{sc})^2 \left(\frac{k'}{k_{F}}\right)^{2l} }} 
+\int_{k_F}^{\infty}  dk' 
\frac{ k' \left( \frac{k'}{k_F} \right)^{-2l} }{\sqrt{(\frac{k'^2}{2m}-\frac{k_F^2}{2m})^2 + (\Delta_F^{sc})^2 \left(\frac{k'}{k_{F}}\right)^{-2l} }} 
\bigg]
\end{equation}
In order to perform the numerical calculation, we need to make the variables dimensionless. Let $x=\frac{k'}{k_F}$, we have
\begin{equation}
\Delta_F^{sc} 
= 
\frac{ \Delta_F^{sc} k_F^2 }{2m} \left[
\int_0^{1}  dx 
\frac{ x^{2l+1} }{\sqrt{(\frac{k_F^2}{2m}x^2-\frac{k_F^2}{2m})^2 + (\Delta_F^{sc})^2 x^{2l} }} 
+\int_{1}^{\infty}  dx 
\frac{ x^{1-2l} }{\sqrt{(\frac{k_F^2}{2m}x^2-\frac{k_F^2}{2m})^2 + (\Delta_F^{sc})^2 x^{-2l} }} 
\right]
\end{equation}
Dividing both sides by $\epsilon_F$ and letting $u \equiv \frac{\Delta_F^{sc}}{\epsilon_F} = \frac{2m\Delta_F^{sc}}{k_F^2}$, we get
\begin{equation}
u = 
\int_0^{1}  dx 
\frac{ u x^{2l+1} }{\sqrt{(x^2-1)^2 + u^2 x^{2l} }} 
+\int_{1}^{\infty}  dx 
\frac{ u x^{1-2l} }{\sqrt{(x^2-1)^2 + u^2 x^{-2l} }}
\end{equation}
We can further simplify the equation by setting $y=x^2$ and obtain
\begin{eqnarray}
u &=& 
\frac{1}{2} \int_0^{1}  dy 
\frac{ u y^{l} }{\sqrt{(y-1)^2 + u^2 y^{l} }} 
+\frac{1}{2} \int_{1}^{\infty}  dy 
\frac{ u y^{-l} }{\sqrt{(y-1)^2 + u^2 y^{-l} }}
\label{Eq:gap2}
\end{eqnarray}
Eq.(\ref{Eq:gap2}) can be solved by iteration using Mathematica,
\begin{eqnarray}
u_{out} &=& 
\frac{1}{2} \int_0^{1}  dy 
\frac{ u_{in} y^{l} }{\sqrt{(y-1)^2 + u_{in}^2 y^{l} }} 
+\frac{1}{2} \int_{1}^{\infty}  dy 
\frac{ u_{in} y^{-l} }{\sqrt{(y-1)^2 + u_{in}^2 y^{-l} }}
\label{Eq:gap3}
\end{eqnarray}

For the singlet spin state, only the even $l$ pairing is allowed, so we have the singlet $d$ wave paring. The superconducting gap will be 
\begin{eqnarray}
\Delta_k^{sc} = |\Delta_k^{sc}| e^{2i\phi_k},
\end{eqnarray}
where 
\begin{eqnarray}
|\Delta_k^{sc}| &=& \left\{ 
\begin{array}{ll}
0.406 \epsilon_F \left(\frac{k_{F}}{k}\right)^{2} & \textrm{ as } k \ge k_{F}, \\
0.406 \epsilon_F \left(\frac{k}{k_{F}}\right)^{2} & \textrm{ as } k \le k_{F},
\end{array}
\right.
\end{eqnarray}
As $k \le k_F$, the gap is
\begin{eqnarray}
\Delta_k^{sc} &\propto& k^2 \left( \cos 2 \phi_k +i \sin 2 \phi_k \right) \nonumber\\
&=& \left( k_x^2 - k_y^2 \right) + 2 i k_x k_y,
\end{eqnarray}
so it will be $(d_{x^2-y^2} + i d_{xy})$ superconducting state.

At finite temperatures, the gap equation becomes
\begin{equation}
\Delta_k^{sc}(T) = - \sum_{k'} V_{kk'} 
\frac{\Delta_{k'}^{sc}(T)}{2\sqrt{\xi_{k'}^2 + |\Delta_{k'}^{sc}(T)|^2 }} 
\tanh \left(
\frac{1}{2k_{B} T} \sqrt{\xi_{k'}^2 + |\Delta_{k'}^{sc}(T)|^2 }
\right)
\end{equation}
Similar to the zero temperature case, the solution for $\ell$ wave pairing will be
$
\Delta_k^{sc}(T) = |\Delta_k^{sc}(T)| e^{i \ell \phi_k},
$
and the magnitude of the gap is now
\begin{eqnarray}
|\Delta_k^{sc}(T)| &=& \left\{
\begin{array}{l} 
\Delta_{F}^{sc}(T) \left(\frac{k}{k_{F}}\right)^{\ell}, \textrm{~~~for~} k\le k_{F} \\
\Delta_{F}^{sc}(T) \left(\frac{k_{F}}{k}\right)^{\ell}, \textrm{~~~for~} k\ge k_{F}
\end{array} \right.
\end{eqnarray}
with temperature-dependent gap $\Delta_{F}^{sc}(T)=|\Delta_{k_{F}}^{sc}(T)|$.

\begin{figure}[htb]
\centering
\includegraphics[width=0.5\textwidth]{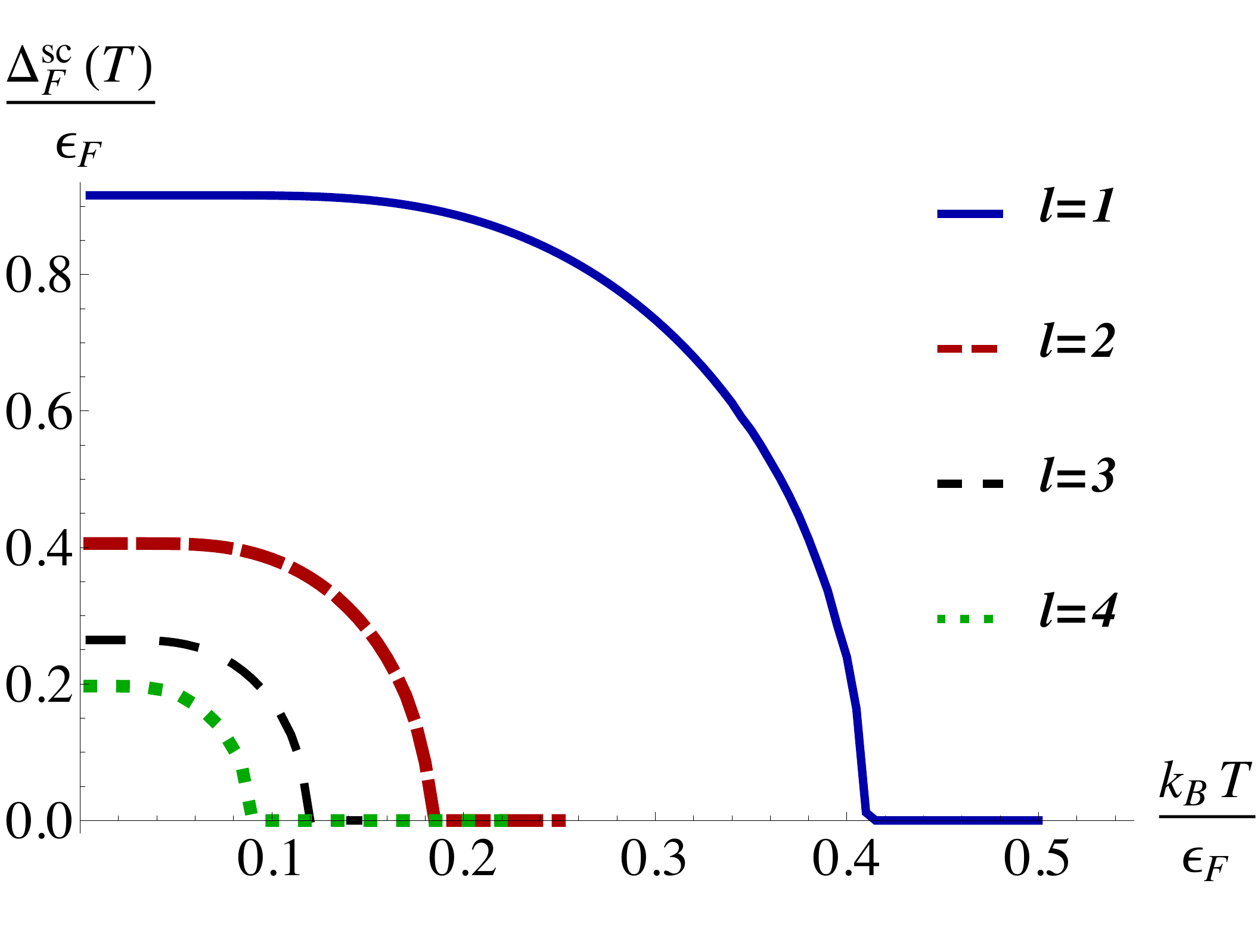}
\caption{Tempertaure dependence of the pairing gap for the angular momentum channels $\ell=1,2,3,4$ from top to bottom. Only $\ell=2,4$ are relevant for singlet pairing.}
\label{Fig:T-dep}
\end{figure}

At a given temperature, the gap equation can be solved by iteration as above. The temperature dependence of the gap $\Delta_{F}^{sc}(T)$ can be found for all non-zero $\ell$; here we show some dominant channels, $\ell=1, 2, 3, 4$, in Figure~\ref{Fig:T-dep}. The zero temperature gap and superconducting transition temperature are listed in Table~\ref{Tab:gap}.
From Table~\ref{Tab:gap} one can see that the ratio $2\Delta_{F}^{sc}(T=0)/(k_B T_{c}) \sim 4.390$ for $\ell=2$, which is comparable to the experimental value 4.9 from the point contact spectroscopy.~\cite{Morales:2009} Notice that both the theoretical and experimental values are larger than the BCS value $\sim 3.52$. 
\begin{table}[htb]
\tbl{The zero temperature gaps and transition temperatures for the angular momentum channels, $\ell=1,2,3,4$.}
	{\begin{tabular}{|c|c|c|c|c|}
	\hline
	$\ell$ & 1 & 2 & 3 & 4  \\
	\hline 
	$\Delta_{F}^{sc}(T=0)/\epsilon_{F}$ & 0.916 & 0.406 & 0.264 & 0.197 \\
	\hline
	$k_B T_{c}/\epsilon_{F}$ & 0.413 & 0.185 & 0.121 & 0.090 \\
	\hline
	$2\Delta_{F}^{sc}(T=0)/k_B T_{c}$ & 4.436 & 4.390 & 4.364 & 4.378 \\ 
	\hline
	\end{tabular}}
\label{Tab:gap}
\end{table}

\section{Discussion}

The purpose of this review was  to explore the role of skyrmions in condensed matter systems and to point out how it can lead to a novel chiral superconducting state via fractionalization into merons and antimerons. Although the framework was cast in the context of HO state in a heavy fermion model, it is the chiral-$d$ wave superconductivity that was the emphasis. 

We have shown that the spin texture in the st-DDW  leads to skyrmions, which can then fractionalize. This is not possible without the mixing of the singlet and the triplet components of the $d$-density wave. The skyrmions acquire charge $2e$ as a result of spectral flow induced by threading flux through the center of the system. Our work should be  interesting because it has been pointed out that, based on the charge, the thermal transport and the specific heat measurements that the superconducting state in the URu$_2$Si$_2$  may have the form $\Delta_k^{sc} \propto k_z (k_x \pm i k_y)$~\cite{Kasahara:2007,Yano:2008}, which is also a chiral $d$-wave superconductor. A more detailed comparison with experiments is beyond the scope of this review. As remarked in the Introduction, we have purposely avoided various complications to get to the essence of the skyrmion physics. However, we are greatly encouraged by the beautiful PKE measurements by Schemm {\em et al.}~\cite{Schemm:2015}, which directly probes breaking of breaking of TRS in the superconducting state.

However, there are a number of unresolved issues:
\begin{enumerate}
\item In the PKE measurements, there is a pronounced anomaly at $T^{*} \sim 0.8-1$K within the superconducting phase.~\cite{Schemm:2015} This may involve the $\ell=4$ channels below the temperature $T_{c}^{(\ell=4)}$. This explanation may be supported by the fact that the ratio of the superconducting transition temperatures for $\ell=4$ and $\ell=2$ channels, $T_{c}^{(\ell=4)}/T_{c}^{(\ell=2)} \sim 0.486$, is comparable to the experimental ratio of the anomaly temperature to the superconducting transition temperature $T^{*}/T_{c} \sim 0.533 - 0.667$. It is tempting to suggest that the subdominant order in $\ell=4$ channel is excited by the large  laser frequency. This could be very similar to $\mathrm{^{3}He}$ where the subdominant pairing in the $f$-wave channel is visible  only in  collective mode measurements.~\cite{Davis:2008,Sauls:1986} It would be interesting to vary the laser frequency, if possible.

 \item A less direct measurement of broken TRS in the superconducting state was recently presented in Ref.~\citenum{Li:2013}. On the other hand, a NMR experiment~\cite{Takagi:2012} finds conflicting results of  broken TRS in the HO state itself. There appear to be no data below $5 K$. So, we do not know how the NMR signature of the broken TRS found in the HO state  relates to that below the superconducting transition temperature below $\sim 1.5 K$.
observed in PKE. Clearly further NMR experiments will be helpful to settle this issue; see also Ref.~\refcite{Schemm:2015}  for PKE at higher temperatures and its interpretation.
\item One of the experimental signatures of the HO state is a specific heat  jump  
$\Delta C/T \approx 270$ mJ/mol-K$^2$ at $T_{\textrm{HO}}$, followed by an exponential behavior below $T_{\textrm{HO}}$, which can be fitted with a gap of $\approx$ 11 meV~\cite{Palstra:1985}. In Fig.~\ref{Fig:specific-heat}, the numerical calculation of the specific heat of the st-DDW state is presented, where the superconducting state is not included. Here an exponential behavior can be seen for $T<T_{\textrm{HO}}$, consistent with  experiments. The chosen  parameters were
$W_{0} = 14$ meV and $\Delta_{0} = 13$ meV. This choice of
the parameters  results in an overall  st-DDW gap of 11 meV; see Eq.~\ref{eq:gap}. The specific heat jump $\Delta C/T\approx 344$ mJ/mol-K$^2$ at $T_{\textrm{HO}}$. Note that we did not include the temperature dependence of the  st-DDW gap, so the results in the region $T \lessapprox T_{\textrm{HO}}$  are overestimated, but this does not change our principal conclusions. We also do not know how to properly model the low carrier concentration semimetallic behavior at low temperatures. It may well be that the linear specific heat is due to two-level systems,~\cite{Anderson:1972} which can be tested by improving the sample quality. There could also be alternate explanations in terms of thermal excitations of merons. A more pedestrian reason could be that the three-dimensional Fermi surface is only partially gapped. This can be seen if we rotate to the diagonal planes, as in our previous work.~\cite{Hsu:2014} We have purposely avoided this rotation in the present review so as not to obscure the primary mechanism.
\begin{figure}[htbp]
\centering
\includegraphics[width=0.5\textwidth]{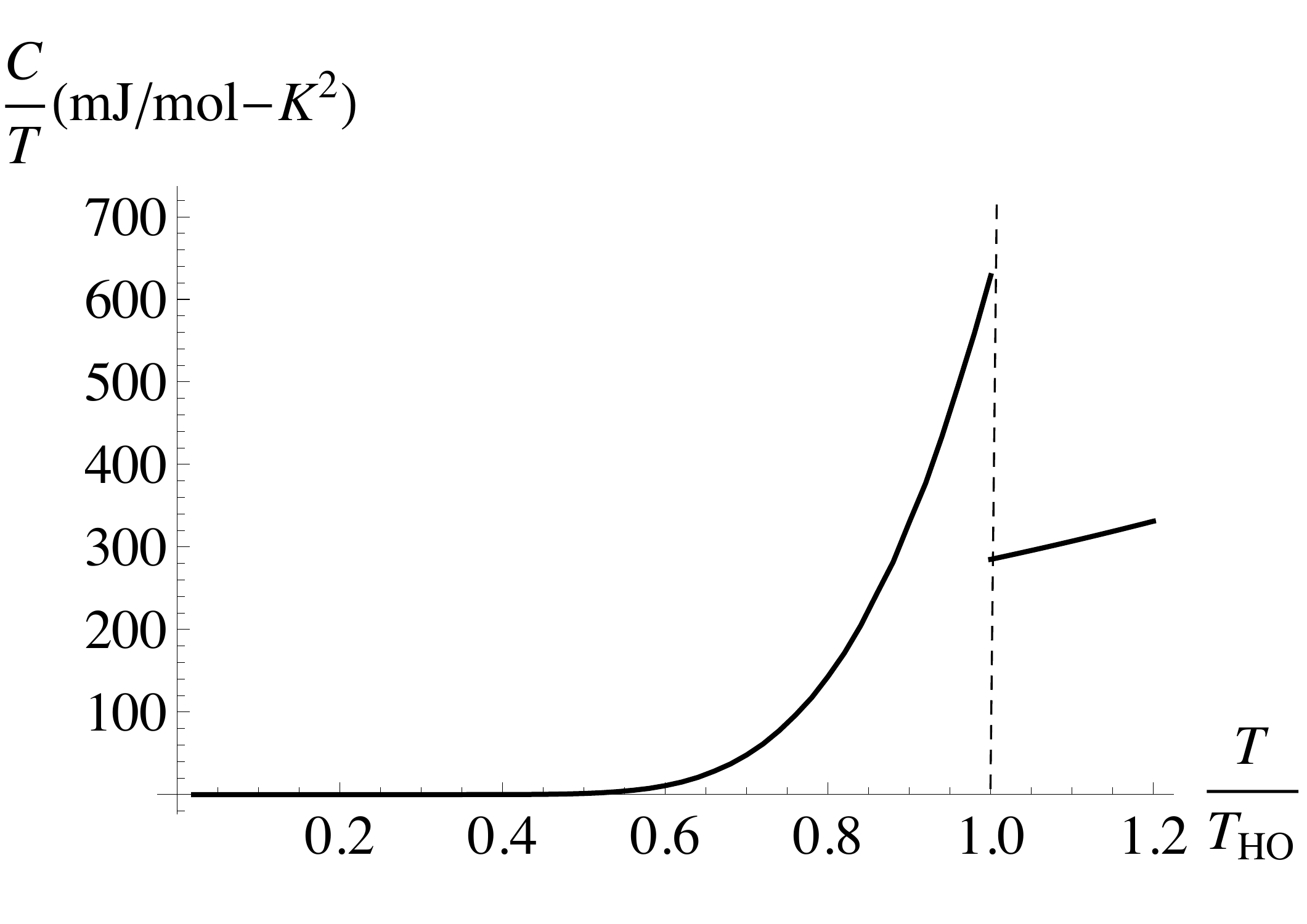}
\caption{The calculated specific heat of the st-DDW state plotted as $C/T$ vs $T/T_{\textrm{HO}}$. The line above $T_{\textrm{HO}}$ is obtained from the experimental data for URu$_2$Si$_2$~\cite{Palstra:1985}.}
\label{Fig:specific-heat}
\end{figure}

\item We assumed the existence of  deconfined merons and antimerons at a deconfined quantum critical point for which it was necessary that hedgehog configurations were suppressed.~\cite{Senthil:2004,Senthil:2004b,Grover:2008} However, we do not know if such a deconfined quantum criticality actually happens, although it results in an exciting possibility. As emphasized above, a bound meron-antimeron pair which forms a skyrmion of charge $2e$ and its subsequent BEC condensation is clearly a different mechanism from the superconductivity from the BCS condensation of merons. If so,  there could in principle be  a BCS-BEC crossover or a transition between the two. It would be interesting to explore this avenue.

\item There may be other orders competing with the superconducting state. One possibility is a Skyrmion lattice, which may be understood by considering the repulsive interaction between the Skyrmions. It will be similar to the Abrikosov vortex lattice due to the vortex-vortex repulsion. If such an interaction exists, the skyrmion lattice  can compete with the superconductivity. However, we do not know how to formulate this problem. Maybe a more general question is whether there are any other competing or coexisting orders due to the non-trivial spin texture.. 
The crystallization of skyrmions has recently been confirmed in neutron scattering studies of the three-dimensional helical magnets MnSi~\cite{Muhlbauer:2009} and Fe$_{1-x}$Co$_x$Si~\cite{Munzer:2010}. The real space imaging of a skyrmion lattice has also been reported in the chiral magnet Fe$_{0.5}$Co$_{0.5}$Si~\cite{Yu:2010}.

\item We assume the length scale of a meron is much smaller than the distance between two merons, so the merons can be treated as point particles; i.e. in the dilute meron approximation. What would happen if we remove  this restriction? The stability of  the superconducting state with respect to Coulomb repulsion between merons must be seriously addressed. The pairing instability should be examined  including the Coulomb repulsion, 
\begin{equation}
\mathcal{H}_{C} = \sum_q V_{q} \rho_{-q} \rho_{q},
\end{equation} 
where $V_{q} = \frac{2\pi}{\epsilon_{d} q}$ is the two dimensional Coulomb potential with the dielectric constant $\epsilon_{d}$.  
However, we do not know how the stability against the meron-meron repulsion relates to the Skyrmion-Skyrmion repulsion which leads to the Skyrmion lattice.

\item In the continuum approximation, we assumed that  the kinetic energy of a meron is $\epsilon_k = \frac{k^2}{2m}$, but we did not justify it. Although, this is the simplest possible assumption, it needs  justification, including the definition of the mass of a meron. Perhaps a more general question is whether the results will hold if we assume a different form for the kinetic energy of the merons. What is the extension to the lattice version?

\item When we constructed the effective Hamiltonian for merons, we set a chemical potential $\mu$ for them. How is this $\mu$ related to the chemical potential in the original electron system? Notice that in the original electron system the chemical potential needs to lie in the gap so that the adiabatic argument for  charge-$2e$ skyrmions  holds. A speculative answer to this question is that the meron chemical potential is half the skyrmion gap, easily obtained from the non-linear $\sigma$-model. Alternately, the meron chemical potential can be inferred from the Fermi energy determined by comparing with experiments.

\item An extremely interesting question is the role of the Goldstones, as inferred from the non-linear $\sigma$-model. Are they merely spectators or do they have important physical consequences? The coupling of the gapless edge modes and the gapless Goldstone modes in the QSHI is an open question.~\cite{Raghu:2008}
 
\end{enumerate}

\section*{Acknowledgements}

We thank E. Abrahams, S. Kivelson, S. Raghu and Z. Wang for discussion. We are particularly grateful to E. Schemm and A. Kapitulnik for keeping us updated about their PKE measurements. This work was supported by NSF under Grant No. DMR-1004520.

\section*{References}


\end{document}